\documentclass[10pt,journal,compsoc]{IEEEtran}
\usepackage[T1]{fontenc}
\usepackage{graphicx}
\usepackage{xspace}
\usepackage{lipsum}
\usepackage{amsmath}
\usepackage{multirow}
\usepackage{tikz}
\usepackage{amsmath,amssymb,amsfonts}
\usepackage{algorithmic}
\usepackage{textcomp}
\usepackage{xcolor}
\usepackage{subfigure}
\usepackage{comment}
\usepackage{algorithm}
\usepackage{pdflscape}
\usepackage{url}
\usepackage[hidelinks]{hyperref}

\newtheorem{definition}{Definition}
\begin{document}

\title{MBTree: Detecting Encryption RATs Communication Using Malicious Behavior Tree }

\author{Cong~Dong, Zhigang~Lu\IEEEauthorrefmark{1}, Zelin~Cui, Baoxu~Liu, Kai~Chen
\thanks{Cong Dong, Zhigang Lu, Zelin Cui, Baoxu Liu and Kai Chen are with Institute of Information Engineering, Chinese Academy of Sciences, and School of Cyber Security, University of Chinese Academy of Sciences, e-mail: (\{dongcong, luzhigang, cuizelin, liubaoxu, chenkai\}@iie.ac.cn).}%
\thanks{This paper is accepted by IEEE Transactions on Information Forensics and Security (TIFS). Due to IEEE copyright limitations, this is the accepted version.}
}

\markboth{Journal of \LaTeX\ Class Files,~Vol.~14, No.~8, August~2015}%
{Shell \MakeLowercase{\textit{et al.}}: Bare Demo of IEEEtran.cls for IEEE Journals}

\IEEEtitleabstractindextext{

\begin{abstract}
Network trace signature matching is one reliable approach to detect active \emph{Remote Control Trojan, (RAT)}. Compared to statistical-based detection of malicious network traces in the face of known RATs, the signature-based method can achieve more stable performance and thus more reliability. However, with the development of encrypted technologies and disguise tricks, current methods suffer inaccurate signature descriptions and inflexible matching mechanisms. In this paper, we propose to tackle above problems by presenting \emph{MBTree}, an approach to detect encryption RATs Command and Control (C\&C) communication based on host-level network trace behavior. MBTree first models the RAT network behaviors as the malicious set by automatically building the \emph{multiple level tree, MLTree} from distinctive network traces of each sample. Then, MBTree employs a detection algorithm to detect malicious network traces that are \emph{similar} to any MLTrees in the malicious set. To illustrate the effectiveness of our proposed method, we adopt theoretical analysis of MBTree from the probability perspective. In addition, we have implemented MBTree to evaluate it on five datasets which are reorganized in a sophisticated manner for comprehensive assessment. The experimental results demonstrate the accurate and robust of MBTree, especially in the face of new emerging benign applications.

 \end{abstract}

}

\maketitle
\IEEEdisplaynontitleabstractindextext
\IEEEpeerreviewmaketitle

\section{Introduction}

\IEEEPARstart{N}{owadays}, modern network attacks are accomplished with advanced automatic tools. One of them used in the post-penetration procedure is the RAT. RAT is essentially a type of software providing convenience for adversaries to complete post-penetration. Typical functions of RAT include process monitor, command execute, keyboard logger, file transfer and others. After break into the victim machine, adversies rely on the sophisticated tool for further steps. Hence, there is a need to specify the malicious activities of RAT to cut off the whole kill chain and prevent potential losses. In the early stage of the RAT development, signature-based methods can accurately detect malicious network activities with distinctive string patterns.
However, since traffic encryption is widely used in different malware including RAT to acquire longer life cycles, the traditional network detection method is greatly challenged.

To tackle the problem brought by encryption technology, novel detection approaches are proposed. Generally, the art methods can be distributed into two main categories according to the detection technology. 
1) Statistical-based \cite{xie2019method, gezer2019flow, wu2019session, li2019method, stergiopoulos2019using, lashkari2017towards, wang2019mobile,kim2019intrusion, huang2018scalable}, which relies on the machine learning or deep learning to learn the classification boundary among different traffic types. It follows the procedure that first produces traffic features like bytes count or interval gap means of packets, and then applies models to train and predict based on the distilled information.
2) Signature-based \cite{papadogiannaki2018otter, trimananda2019pingpong}, which relies on abstract behaviors of the encrypted traffic used as identification signatures. It first formulates distinctive elements of the traffic as the signature set, and then check if the monitored traffic implies any appointed signature.

In recent years, the computer security community mainly follows the first technical route to utilize the powerful learning ability of machine learning models to tackle the problem. However, this approach is not perfect. 
First and most important, statistical models perform unstable in different environments. The learning strategy of most machine learning models requires not only malicious traffic but also benign traffic in training procedure \cite{li2018machine}. Since the benign applications vary in different environments, the model will be confused by the unknown application network traces not appeared in the training procedure; thus the trained model will perform unstable when transferred into a different environment.
Second, the method also requires sufficient labeled instances in the training procedure. To meet this requirement, a large amount of data should be collected in advance, which dramatically increases the inconvenience of applying the method in a real environment. Considering the shortage of statistical-based approaches, hence, we focus on signature-based routine. Compared to the former technic, the signature-based approaches can achieve robust performance through different environments and thus more reliable in a real application. Besides, the signature-based approach requires only a few malicious network traces, which can release the tedious work of data gathering.

\subsection{Problems}
Traditional signatures, like string patterns, can hardly adapt to encryption context because the payload is unknown during data transmission. Motivated by the observation that applications usually follow a fixed-code procedure to initiate or respond to requests, the snippets of \emph{directed packet payload size (DirPiz)} sequence of a flow in arbitary position is used by several arts as fingerprints to identify encrypted behaviors \cite{trimananda2019pingpong, papadogiannaki2018otter}. 
However, these proposed methods are not able to detect RAT in an accurate manner. 
On one hand, false positives can occur due to the Dirpiz conflicts. Potentially, the snippets of flow-level DirPiz sequences generated from a benign application can be the same as that from RATs, thereby causing false alarms. In another word, only flow-level DirPiz snippets are not precise enough as malicious signatures. 
On the other hand, false negatives can occur because the DirPiz varies slightly from different environments due to \emph{dynamically generated packets}. Generally, the size of packets carrying the instruction which asks the trapped machine to report the base information is the same, yet the size of corresponding callback packets is different. Unfortunately, existing arts can not properly handle the dynamic packets because all of them rely on the exact matching mechanism of the regex engine.

\subsection{Our Work}
In this paper, we present \emph{Malicious traces Behavior Tree (MBTree)}, a novel signature-based approach for robust and accurate encrypted C\&C identification based on network behavior. 
First, MBTree creates signatures to depict malicious behaviors by integrating flow-level DirPiz sequences as a synthesis of host-level \emph{Multi-Level Tree (MLTree)} to improve the unique degree of malicious descriptions. In such a way, approximately 94\% false alarms triggered by only a few coincidences DirPiz can be reduced. 
Second, MBTree relies on a flexible similarity-based matching mechanism expanding the coverage of each signature to facilitate robust detection. The proposed mechanism can cover reasonable deviations from the generated signatures. Besides, the alarm level of the detection can be adjusted by a predefined threshold. 

To demonstrate the effectiveness of our approach, we conduct solid experiments on several datasets. These datasets are reorganized in a sophisticated manner to simulate the situation that the test set contains \emph{unknown applications} in a real environment. To acquire stable results, we also integrate the 5-fold cross-validation strategy. The experiment results show that MBTree yields a more accurate performance than machine learning state-of-the-arts on the test sets. Individually, MBTree can achieve approximately 94\% F1-score on validation set, and 91\% F1-score on test set. Moreover, we also analyze the influence of different hyperparameters of them by tuning them and inspect the malicious behaviors by reviewing generated MLTree signatures.

\subsection{Contributions}
In summary, this paper makes the following contributions:
\begin{itemize}
\item First, a sharp network behavior representation method, MLTree, is proposed as the
enhancement signature. It helps to reduce the probability of DirPiz conflicts by integrating multiple related flow-level DirPiz sequences; thus, it can depict encrypted malicious network traces in a more accurate and robust manner. 
\item Second, a corresponding detection mechanism based on the similarity comparison of MLTree is proposed. This strategy enhances the flexibility of detection by covering deviations slightly different from generated signatures. Compared to the exact matching strategy, it can handle the dynamic packets properly; thus, more robust detection can be achieved. 
\item Third, we demonstrate the effectiveness of our approach from both theoretical and experimental aspects. From theoretical perspective, we illustrate that it is in an extremely low probability that MBTree would misclassify different applications. From experimental perspective, persuasive experiments are conducted on different datasets to evaluate the real performance. 
\end{itemize}

The remainder of this paper is structured as follows. Section II summarizes related works and their limitations. Section III gives an overview of the design of MBTree. Section IV elaborates the details of MLTree. Section V describes the similarity matching mechanism. Section VI provides the theoretical analysis of MBTree. 
Section VII covers the experiment setup description. Section VIII reports the experiment evaluation results and analysis. Section IX discusses potential evading strategies to attack MBTree. And section X provides the conclusion.
\begin{table*}
\centering
\footnotesize
\caption{Comparison of different signature-based methods.}
\label{tab_sig_compare}
\begin{tabular}{c|ccccccc} 
\hline
 \textbf{Method}         & \begin{tabular}[c]{@{}c@{}}\textbf{ Signature}\\\textbf{~Generation} \end{tabular} & \begin{tabular}[c]{@{}c@{}}\textbf{Main Element }\\\textbf{of Signature} \end{tabular} & \textbf{Level}         & \textbf{Type}                & \textbf{Scene}            & \begin{tabular}[c]{@{}c@{}}\textbf{Encrypted }\\\textbf{Detection} \end{tabular} & \begin{tabular}[c]{@{}c@{}}\textbf{Detection }\\\textbf{Engine} \end{tabular}  \\ 
\hline
Snort                    & Manual                                                                             & String Pattern                                                                         & packet/flow            & Exact                        & IDS                       & No                                                                               & Self                                                                           \\
Suricata                 & Manual                                                                             & String Pattern                                                                         & packet/flow            & Exact                        & IDS                       & No                                                                               & Self                                                                           \\
Zeek(Bro)                & Manual                                                                             & String Pattern                                                                         & packet/flow            & Exact                        & IDS                       & No                                                                               & Self                                                                           \\
                         & Automatic                                                                          & String Pattern                                                                         & packet/flow            & Exact                        & Malware                   & No                                                                               & Snort                                                                          \\
                         & Semi-Automatic                                                                     & String Pattern                                                                         & flow                   & Exact                        & Malware                   & No                                                                               & Snort                                                                          \\
OTTer                    & Automatic                                                                          & Packet Length                                                                          & flow                   & Exact                        & Application               & Yes                                                                              & Self                                                                           \\
PingPong                 & Automatic                                                                          & Packet Length                                                                          & packet pair            & Exact                        & IOT Application           & Yes                                                                              & Self                                                                           \\
\emph{\underline{MBTree}}  & \emph{\underline{Automatic}}                                                         & \emph{\underline{Packet Length}}                                                         & \emph{\underline{host}}  & \emph{\underline{Similarity}}  & \emph{\underline{Malware}}  & \emph{\underline{Yes}}                                                             & \emph{\underline{Self}}                                                          \\
\hline
\end{tabular}
\end{table*}
\begin{figure*}[htb]
\centering
\centerline{\includegraphics[width = 0.83\linewidth]{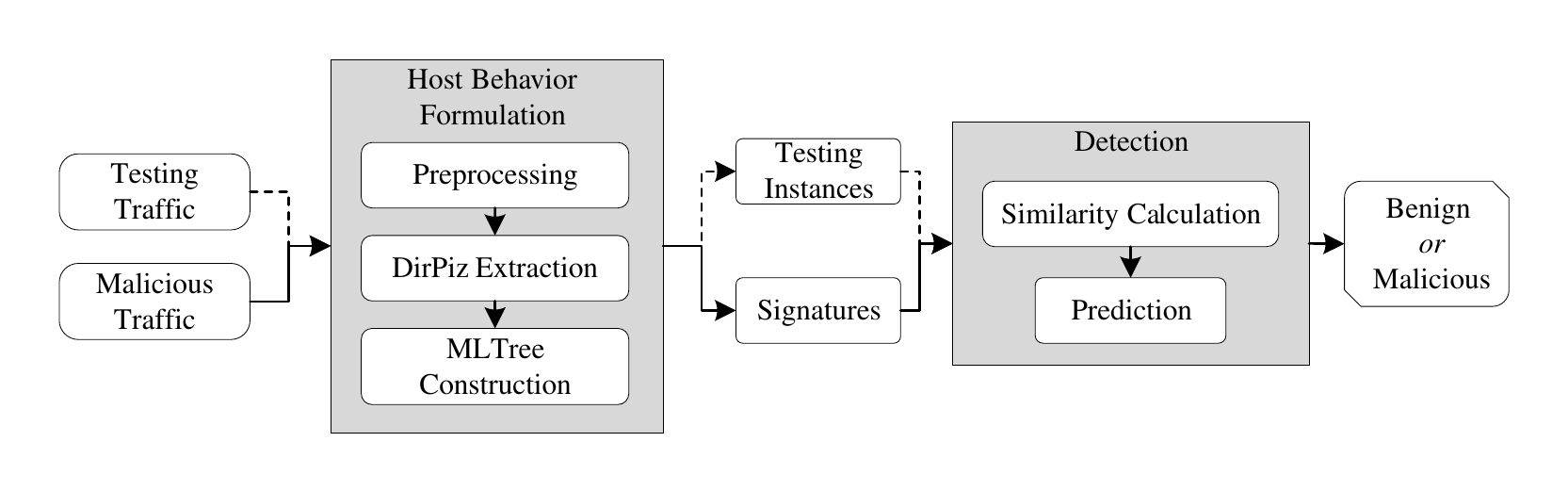}}
\caption{High-level overview of MBTree}
\label{fig_overview}
\end{figure*}
\section{Related Work} \label{sec:relatedwork}
MBTree makes contributions to the problem of encryption RAT C\&C traffic detection. A central idea behind MBTree is adopting the MLTree as signatures. Below, we discuss related works in the above areas from two perspectives.

\subsection{Statistical-based} 
In recent decades, statistical-based methods for RAT traffic detection have been widely studied \cite{xie2019method, gezer2019flow, wu2019session, li2019method, stergiopoulos2019using, lashkari2017towards, wang2019mobile,kim2019intrusion, huang2018scalable, alshammari2011can, sun2010novel, draper2016characterization, casino2019hedge, shen2017classification, kwon2016catching}. Generally, most of these methods follow such procedures that first extract features from the traffic, and then using statistical models to fit the data. 

Recently, \cite{gezer2019flow, stergiopoulos2018automatic, stergiopoulos2019using, kim2019intrusion} use side-channel information with Random Forest (RF) for encrypted malicious traffic detection. Specifically, these side-channel features range from packet length, packet interval time to payload ratio. The experiment results have shown the effectiveness of this combination of side-channel features and machine learning models. 
Besides, advanced deep learning models are also adopted in this area \cite{li2019method, xie2019method, aceto2019mimetic, lotfollahi2020deep, aceto2019mobile, dong2020cetanalytics}. Compared with traditional machine learning models, these models do not require the feature extraction procedure. The sophisticated models can automatically extract related features, and achieve end-to-end classification. For example, 
\cite{li2019method} proposes a deep learning method to detect HTTP malicious traffic on mobile networks. 
\cite{xie2019method} proposes a sophisticated deep learning architecture to detect trojans with hierarchy spatiotemporal features. 
Generally, these traditional machine learning methods and advanced deep learning methods can achieve high performance in their experiments. However, they can hardly adapt to the situation of unseen applications beyond the training set due to the inconsistent statistical distribution. Moreover, they both lack the interpretability to trace back the cause of alarms for further response. 

\subsection{Signature-based}
Different from statistical-based methods, signature-based methods rely on previously defined elements of the traffic, mainly on specific content string patterns. Generally, most of them first create a signature set and then determine if the testing instances match one or more signatures in the set. 
Compared with machine learning methods, signature-based methods are more robust among different environments; thus, it is still applied in multiple security products, ranging from intrusion detection systems (IDS) to firewalls \cite{d2019survey}.

Despite the advance, there exist two apparent limitations of traditional signature methods. First, most methods require manual work to produce high-quality signatures even though given the cleaned malicious traffic. This manual work is tedious for security analysts. Second, traditional signatures like specific strings are seriously challenged by encrypted traffic \cite{roesch1999snort,paxson1999bro,suricata}. Because the string patterns are invalid to ciphertext.
Several studies focus on the automatic generation of network signatures to tackle the first problem \cite{newsome2005polygraph, perdisci2010behavioral, tegeler2012botfinder, shim2017sigbox}. For example, 
\cite{newsome2005polygraph} propose three automatically generated signatures ranging from conjunction signatures, token-subsequence signatures to Bayes signatures for polymorphic worm detection with traffic payload.  
\cite{perdisci2010behavioral} uses clustered malicious traces to produce network-level signatures from HTTP fields automatically. These network-level signatures are translated in a format compatible with Snort rules and can be used for detection.
Apart from only focusing on the first problem, recent studies focus on tackling both problems by introducing novel behavior patterns. Instead of using specific strings as signatures, these novel methods use other aspect information to identify specific behaviors, like packet size sequence. For example, \cite{papadogiannaki2018otter} propose an encrypted traffic pattern language based on \emph{packet payload size sequence} for scalable Over-The-Top (OTT) applications identification. They show that this signature is unique to identify applications or even application events.
Similarly, \cite{trimananda2019pingpong} proposes a method based on packet length pairs to identify specific events of home IoT devices. To achieve the fine-grained event-level traffic detection, \cite{trimananda2019pingpong} adopt the DBSCAN algorithm to search frequent conversation pairs, and then concatenate the pairs into sequences as signatures. 

Even though previous studies present the effectiveness of novel network signatures, they still lack the ability to detect encrypted RAT in an accurate and robust manner. This can be attributed to the inaccurate signature building and inflexible matching mechanism. Since several RATs can establish different connections to evade the detection \cite{antonakakis2017understanding, lever2017lustrum}, only flow-level fingerprints can hardly cover the whole picture of the encryption RAT behaviors. Besides, although the exact matching strategy can cooperate with the finite state machine to improve efficiency, it lacks the flexibility to capture similar behaviors among different environments. To clearly show the path of research in this direction, we summarize the differences of each method in \autoref{tab_sig_compare}.

\section{MBTree Overview}
\begin{figure*}[htb]
\centerline{\includegraphics[width=0.73\linewidth]{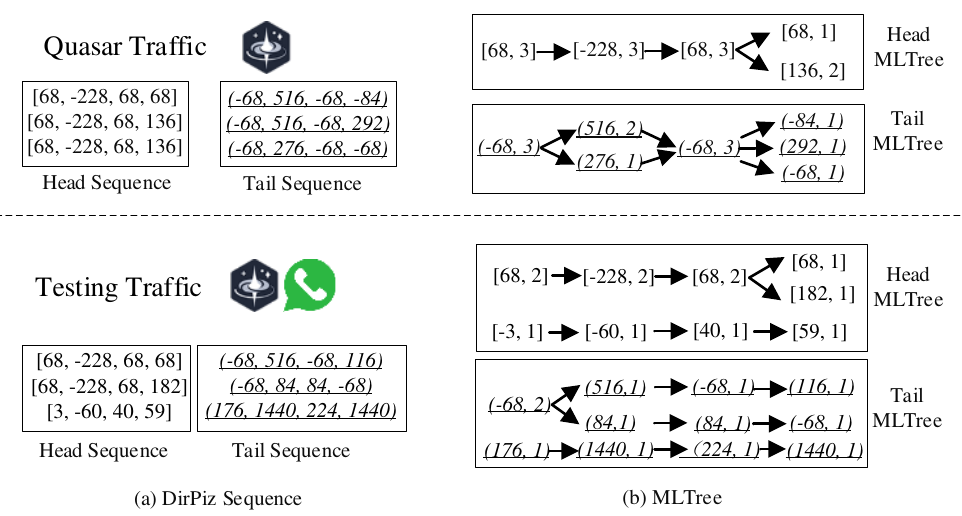}}
\caption{A brief example shows the simplified DirPiz sequences of each session with setting $L$ as 4. The training set contains the malicious traffic generated from Quasar. The testing set contains the mixed traffic generated from Quasar and WhatsApp. Positive numbers denote the packet sent from the client, while negative numbers denote the packet sent from the server. 
}
\label{Fig_total_MLTree}
\end{figure*}
In this section, we provide an overview of the proposed MBTree system. As shown in \autoref{fig_overview}, our approach consists of two main procedures, \emph{host behavior formulation} and \emph{detection}. First, the raw traffic is formulated to MLTree representing host-level behaviors in three steps. 1). The preprocessing achieves traffic cleaning and session reassembling. 2). The DirPiz sequences from each session as multiple independent fingerprints. 3). The sequences are correlated based on common hosts to construct MLTree as the host behavior. It is worth mentioning that malicious signatures and testing instances are produced following the same steps in this procedure.
Second, the converted testing instances are compared to each signature to decide if they match any malicious behaviors in two steps. 1). The similarity between the testing instance and each signature is calculated to produce a similarity vector. 2) whether the instance belongs to malicious is predicted based on the similarity vector.

To show the workflow clearly, we also provide a simplified example throughout the following sections to describe the key design of MBTree. The example uses a part of pure Quasar traffic as malicious traffic, and a part of mixed Quasar traffic and WhatsApp traffic as testing traffic, as shown in \autoref{Fig_total_MLTree}.

\section{Host Behavior Formulation}
In this section, we describe the details of the formulation procedure from raw traffic to the host-level signature MLTree. \autoref{sec:preprocessing} introduces the cleaning and session reassemble strategy, \autoref{sec:dirpizextraction} shows our consideration of DirPiz extraction, and \autoref{sec:mltreeconstruction} provides the definition of MLTree and corresponding construction process.

\subsection{Preprocessing} \label{sec:preprocessing}
Raw traffic is usually captured in pcap or pcapng format recording a lot of communication details. Extra meta information should be removed before building the signature. Hence, we first apply traffic cleaning to discard invalid packets, then we apply session reassembling to recover the whole communication.

\textbf{Cleaning;}
Traffic cleaning aims at filtering out redundant information in raw traffic. As the requirements of cleaned traffic for signature and testing instance formulation are different, we apply different strategies for the two sets. For the training set, we apply a `whitelist' strategy to only reserve the packets containing C\&C IPs. For other traffic in the malicious communication, we just discard them. Thus, only ensured communications between the victim and C\&C are selected to produce valid signatures. For the testing set, we apply a `blacklist' strategy to only discard packets that meet the following conditions, repeated packets, loop packets, non-transmission packets. With this strategy, we can reserve the main information of testing traffic and avoid potential malicious packets bypassing.

\textbf{Session reassembling;} After cleaned, the traffic is reassembled into sessions to recover all end-to-end communication. In this procedure, we mainly identify a transmission session based on 5-tuple. The 5-tuple includes source IP, destination IP, source port, destination port, and protocol. Besides, for TCP protocol, the flag fields representing the communication status are also used to identify different sessions using the same 5-tuple.

\subsection{DirPiz Sequence Extraction}\label{sec:dirpizextraction}

Given a reassembled session, we extract the DirPiz sequence of the session to fingerprint the automated procedure. 
As a matter of fact, multiple types of meta-information of the packet can be extracted as fingerprints, such as packet gap intervals, and packet payload hash. However, they either lack stable performance in different network environments or do not adapt to variable encrypted content because of the dynamic encryption key negotiation mechanism. As a result, we choose the DirPiz as the meta information for its robust performance.
Typically, a DirPiz sequence consists of the payload size of each packet with the direction in a connection. The direction means that the packet is either request from the client to the server or a response from the server to the client. Here we provide an example of extracted DirPiz sequences in Fig. 2(a). The Quasar traffic is used as a training set. While the mixed traffic of Quasar and WhatsApp is used as a testing set.

Then we provide the specific procedures to produce a DirPiz sequence. 
First, different IP packets are reassembled if the packet is fragmented. Since the communication content can be scattered in multiple IP packets due to the limitation of MTU or potential IP fragment attack, reassembling these fragmented IP packets can restore the real payload in a communication.
Second, based on these reassembled IP packets, the payload sizes are counted according to the upper-level protocols to form a sequence. In this step, the influence of low-level protocol details can be reduced, e.g., TCP handshake or SSL/TLS negotiation. Different strategies are applied according to different protocols. For UDP protocol, the length of the transport payload is used directly. For TCP protocol, only the length of the transport payload after the connection is established is used. For SSL/TLS protocol, only the length of the payload in the 'Application Data' packet is used.
Third, we append the direction sign to the elements of payload size sequences. The request information from a client to the server is formulated as a positive number, and the response information from a server to the client is formulated as a negative number. 
Fourth, all sequences are aligned to the same length to keep consistent. 
Only \emph{$L$ DirPiz} are reserved in the sequence. Besides, for sequences less than $L$ in length, 0 is used as padding value. For example, Fig.~2(a) shows the extracted DirPiz sequences with setting $L$ as four.

\subsection{MLTree Construction}\label{sec:mltreeconstruction}
\begin{algorithm}[h]
  \caption{MLTree Initialization}
  \label{alg_mltree_construct}
  \begin{algorithmic}[1]
  \REQUIRE DirPiz Sequence Set $P$, Max Level $L$
  \ENSURE MLTree $T$
  \STATE Let $T = (N, E, C_N, C_E)$
  \STATE Let $N=\varnothing, E=\varnothing,C_N=\varnothing,C_E=\varnothing$
  \FORALL{level ${l} \in L$}
  
  \STATE $N^l = \varnothing, C_N^l=\varnothing, E^l=\varnothing, E_N^l=\varnothing$
  \FORALL{DirPiz sequence ${p} \in P$}
  \STATE $N^l = N^l \cup \{p[l]\}$, $C_N^l[p[l]] +=1$
  \IF {$l$ eq 0}
  \STATE $e=(0, p[l]), E^l = E^l \cup \{e\}$, $C_E^l[e]+=1$
  \ELSE
  \STATE $e=(p[l-1], p[l]), E^l = E^l \cup \{e\}, C_E^l[e]+=1$
  \ENDIF 
  \ENDFOR
  \STATE Append $N^l, C_N^l, E^l, C_E^l$ to $N,C,C_N,C_E$
  \ENDFOR
  \RETURN T
  \end{algorithmic}
\end{algorithm}

In this section, we integrate multiple diverse DirPiz sequences into MLTree as the host signature. 
Corresponde to head sequences and tail sequences, two MLTrees are used to represent the host signature, specifically, a head MLTree and a tail MLTree.  
As a central structure of our approach, MLTree is defined as follows, 

\begin{definition}
\textbf{MLTree} MLTree $T$ is a Weighted Directed Acyclic Graph (WDAG), $T = (N, E, C_N, C_E)$, where $N, E, C_N, C_E$ represent the node set, edge set, node weight set, edge weight set respectively. The node set $N$ is used to represent unique DirPiz grouped by level. The statistical set $C_N$ is used to record the aggregated occurrence of each unique DirPiz organized in level. The edge set $E$ is used to represent two co-occurrence DirPiz grouped by two adjacent levels. The statistical set $C_E$ is used to record the co-occurrences of each two adjacent DirPiz. 
With different to normal WDAG, MLTree is organized in hierarchy structure, every weighted node $n_i^l$ and weighted edge $e_i^l$ belongs to a specific level sub-set, $n_i^l \in N^l, e_i^l \in E^l$. And all these level sub-sets $N^l, E^l, C_N^l, C_E^l$ consist of the corresponding elements, $N=\{N^l\}_{l=0..L},E=\{E^l\}_{l=0..L},C_N=\{C_N^l\}_{l=0..L},C_E=\{C_E^l\}_{l=0..L}$.  
\end{definition}

\begin{algorithm}[h]
  \caption{MLTree Merging}
  \label{alg_mltree_merge}
  \begin{algorithmic}[1]
  \REQUIRE MLTree Set $M$
  \ENSURE Merged MLTree $T$
  \STATE Let $T = (N, E, C_N, C_E)$
  \STATE Let $N=\varnothing, E=\varnothing,C_N=\varnothing,C_E=\varnothing$
  
  \FORALL{level ${l} \in L$}
  \STATE $N^l = \varnothing, C_N^l=\varnothing, E^l=\varnothing, E_N^l=\varnothing$
  \FORALL{MLTree ${m} \in M$}
  \STATE $N^l = N^l \cup m.N^l, E^l= E^l \cup m.E^l$
  
  \FORALL{node $n \in N^l$, edge $e \in E^l$}
  \STATE $C_N^l[n] += m.C_N^l[n]$, $C_E^l[e] += m.C_E^l[e]$
  \ENDFOR
  \ENDFOR
  \STATE Append $N^l, C_N^l, E^l, C_E^l$ to $N,C,C_N,C_E$
  \ENDFOR
  \RETURN T
  \end{algorithmic}
\end{algorithm}

The specific MLTree construction process is as follows.
First, the DirPiz sequences are processed level by level as the initial MLTree. In each level of processing, the unique nodes, edges, and their corresponding statistics are specified. The specific initialization schema is shown in \autoref{alg_mltree_construct}.
Second, merging operations are taken to enhance the coverage as well as reduce the storage cost of repeated patterns. The specific merging algorithm is shown in \autoref{alg_mltree_merge}. Intuitively, it contains \emph{polluting risk} when integrating a large number of individual MLTrees into a huge MLTree. Because the DirPiz coverage of the large tree will increase dramatically resulting in high-level false alarms. We divide this kind of situation into two categories for discussion. 1) All initial MLTrees are from the same RAT. In this situation, all these individual MLTrees denote similar DirPiz coverage, thus they can be merged without worrying about polluting risk. 2) Initial MLTrees are from different RAT. In this situation, there is a high polluting risk in merging these individual MLTrees because their DirPiz coverage is different.
To avoid this, we limit the merging operation to integrate the DirPiz sequences generated from the same malicious sample. Thus, similar behaviors can be merged regardless of the number of DirPiz sequences, and the scale of the tree can be limited to a proper extent.
Example MLTrees built from Fig.~2(a). is shown in Fig~2(b). It is apparent that MLTree is hierarchically organized.

Briefly, MLTree provides several advantages to depict malicious behaviors. 
First, MLTree ensures flexible merging to represent signatures efficiently. In our design, different MLTrees derived from the same malicious sample can be merged into a single MLTree. This feature can enhance the signature update ability as well as reduce the cost of storing repeated signatures.
Second, the hierarchy design can represent the host behavior in an intuitive and reasonable manner. Since the frequent DirPiz is with a higher probability of being generated automatically, the automated handshake behavior can be quantified by differentiating between frequent DirPizs and infrequent DirPizs hierarchically. 
Third, MLTree can be automatically constructed given the cleaned malicious traffic. Both the MLTree construction and merging can be completed with a systematic script. In such a manner, the construction requires no interactions with security experts and prior knowledge. Hence the labor costs can be reduced.

\section {Detection}

Unlike traditional signature-based detection, our detection relies on similarity matching strategy instead of exact matching strategy to decide if the testing instance should be regarded as malicious.
Brief steps of detection are as follows. First, the similarity vector of the testing instance and each signature is calculated as $v_{m} = [s_0, s_1, ..., s_n]$. The element in the vector represents the similarity score of the testing instance and the corresponding signature. The specific similarity score calculation method is elaborated in \autoref{sec:similaritycalculation}. Second, we make a prediction based on the similarity vector to decide if the testing instance contains malicious behavior. The details of this step are shown in \autoref{sec:prediction}.

\begin{figure*}[h]
\centerline{\includegraphics[width=0.65\linewidth]{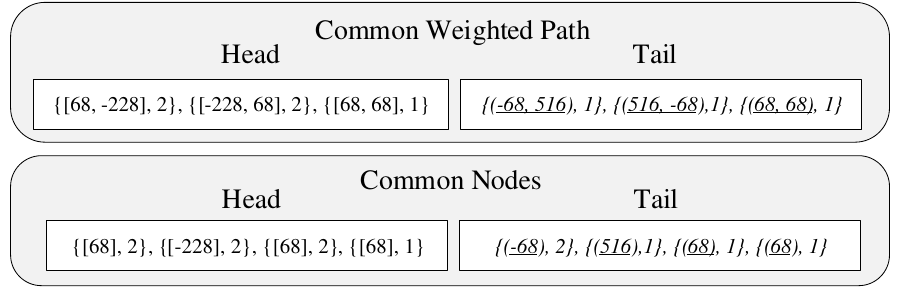}}
\caption{ CWP and Common Nodes of corresponding MLTrees in Fig.~2(b).}
\label{Fig_total_similarity}
\end{figure*}

\subsection{Similarity Calculation} \label{sec:similaritycalculation}
The first step towards malicious prediction is to produce the similarity vector of the testing instance with pre-produced MLTree signatures. In this step, the core is to calculate the similarity of two MLTrees. Roughly, the MLTree similarity is measured from two aspects, \emph{path similarity} and \emph{node similarity}. The path similarity measures the similarity of continuous edges in corresponding hierarchies, while the node similarity measures the similarity of nodes in corresponding hierarchies.
Specifically, a similarity score $S$ of the testing instance and a signature is formulated as
\begin{equation}
\label{eq1}
S = \beta [\alpha S_{fp} + (1-\alpha) S_{fn}] + (1-\beta)[\alpha S_{lp} + (1-\alpha) S_{ln}]
\end{equation}
where $S_{fp}$ and $S_{fn}$ represent the path similarity and the node similarity of corresponding head MLTree respectively, $S_{lp}$ and $S_{ln}$ represent the path similarity and the node similarity of corresponding tail MLTree respectively, $\alpha$ represents the path score ratio parameter that determines the balance of the path score and node score, and $\beta$ represents the head ratio parameter that determines the balance of the head score and tail score. In addition, special DirPiz that appear frequently in both benign and malicious MLTrees can be removed in calculating the similarity score. Thus, identical DirPizs can have a more important influence in similarity calculation.

\subsubsection{Path Similarity}
Path similarity is used to measure the common continuous paths of two MLTrees. Towards achieving this measurement, we first define \emph{common weighted path (CWP)} as follows, 
\begin{definition}
\textbf{Common Weighted Path}
Given two MLTrees $bt = \{N_{t}, E_{t}, C_{Nt}, C_{Et}\}$, $bm = \{N_{m}, E_{m}, C_{Nm}, C_{Em}\}$, a CWP $P_C$ is defined as the intersection of continuous weighted edges of two MBTrees. Specifically, $P_C = \{E_C, C_{Ec}\}$, the edge set $E_I$ is the continuous intersection of the two edge sets $E_t$ and $E_m$. This means $ \forall 1 < l < L$, if $(n_i^l, n_j^{l+1}) \in E_I^l$, then $(n_i^l, n_j^{l+1}) \in (E_{t} \cap E_{m})$, and there exists at least one edge $(n_p^{l-1}, n_i^l)$ that $(n_p^{l-1}, n_i^l) \in E_I^{l-1}$. Besides, the statistics set $C_{Ec}$ based on $E_C$ is also the intersection of the two statistics sets according to different levels. 
\end{definition}

Generally, CWP aims at capturing the common successive sequential information of two MLTrees. Thus, it is used as the middleware to measure the continuous similarity.
The algorithm to generate CWP of two MLTrees is shown in ~\autoref{alg_mbtree_merge}.
\begin{algorithm}[h]
  \caption{CWP Generation}
  \label{alg_mbtree_merge}
  \begin{algorithmic}[1]
  \REQUIRE Testing MLTree $bt = \{N_t, E_t, C_{Nt}, C_{Et}\}$, and signature MLTree $bm = \{N_m, E_m, C_{Nm}, C_{Em}\}$, Max Level $L$. 
  \ENSURE Common weighted path $P_C$
  \STATE Let $P_c = \{E_c=\varnothing, C_{Ec}=\varnothing\}$
 
  \FORALL{level ${l} \in L$}
  	\IF{$l$ is not 0 and $E_c^{l-1}$ is $\varnothing$}
  	\RETURN $P_c$
  	\ENDIF
    \STATE Let $E_c^l=\varnothing, C_{Ec}^l=\varnothing, N_{tmp}^l =N_t^l \cap N_m^l$
    		\IF {$l$ is not 0}
  		\STATE $E_c^l$ = $E_t^l \cap E_m^l \cap (N_{tmp}^l \times N_{tmp}^{l-1})$
  		\ELSE
  		\STATE $E_c^l$ = $E_t^l \cap E_m^l$
  		\ENDIF
  		
  		\FORALL{edge ${e} \in E_c^l$}
  		\STATE $C_{Ec}^l = C_{Em}^l[e] $
  		\ENDFOR
  \ENDFOR
  
  \RETURN $P_c$
  \end{algorithmic}
\end{algorithm}

Then we provide several observations of CWP.
1) Edge statistics of CWP can reflect the similar level of two MLTrees.
2) Range of edge statistics at the same level in different CWP is inconsistent, because the testing instance may contain the traffic generated from a different period than signature traffic.
3) Range of edge statistics at different levels in CWP is inconsistent because the edge statistics are generated independently.
4) It is of low probability to produce a long CWP between two unrelated MLTree. As a matter of fact, two unrelated MLTree may contain the same DirPiz at a level; however, they can hardly contain the same edges or even continuous edges through different levels. Thus, a long CWP can indicate a higher level of similarity than a short CWP.

Next, we synthesize the observations of CWP and provide several considerations to design the formula for path similarity. 
1) A higher similarity score should indicate more similar the two MLTrees are. To achieve this, we design a weighted product mechanism to ensure that the score increases monotonically with the increase of the number of edges and levels.
2) Edge statistics should be normalized to be used as a base factor of the score. Specifically, we use the edge statistics of signatures at the corresponding level to normalize that of testing instance; thus, the impact of the difference among levels can be eliminated. Besides, we also introduce a time ratio to balance the period difference between the testing instance and the signature.
3) Hierarchy level can be used as a weight factor for edge statistics. To leverage the continuous property of CWP, we assign different weights to statistics at different levels. Thus a longer CWP can correspond to higher similarity.
4) the statistics should be normalized to reduce the influence of different time lengths. %

Based on these considerations, we tried several schemes and performed experiments on a small training set. Finally, we found the best result is achieved by the formula as follows, 
\begin{equation}
\label{eq_path_simi}
{{\rm{S}}_p} = {{\rm{2}}^{L' + \sum\limits_{l = 1}^{L'} {[\frac{{F(E_C^l,C_{Em}^l)}}{{F(E_m^l,C_{Em}^l)}} \times \frac{{{l^2}}}{{L'}} \times {R_t}]} }}
\end{equation}
where $L'$ denotes the max level of CWP, ${l^2 \over L'}$ denotes the \emph{level important factor}, the ${\frac{{F(E_C^l,C_{Ec}^l)}}{{F(E_m^l,C_{Em}^l)}}}$ denotes the \emph{normalized level path similarity factor}, $F(a,b)$ represents \emph{counting the total occurrence of all elements of $a$ in set $b$}, $R_t={\frac{{T_m}}{{T_t}}}$ denotes \emph{the time ratio for normalization}, the $T_m$ represents the capturing time of the signature traffic and the $T_t$ represents the capturing time of the testing traffic. It is worth to mention that we add level $L'$ in exponential part as an score normalization factor to differentiate the minum value with different max level $L$.

\subsubsection{Node Similarity}
Node similarity is used to measure the common nodes of two MLTrees in corresponding hierarchies. As preliminary, we define the \emph{Common Nodes} as follows,
\begin{definition}
\textbf{Common Nodes}
The common nodes $N_I$ are defined as the  intersection nodes and their corresponding minimal occurrence at each level, formally, $ I_N = \{N_I, C_{NI}\}, \forall l < L \to N_I^l \subseteq (N_t^l \cap N_m^l), C_{NI}^l \subseteq (C_{Nt}^l \cap C_{Nm}^l)$.
 \end{definition}
 
Totally, the node similarity is calculated based on the \emph{common nodes}.
Unlike CWP, common nodes are generated independently through different levels.
Thus, the nodes are treated equally to contribute to the diverse similarity of two MLTrees.
Specifically, correspond to produce path similarity, the formula to calculate node similarity is as follows,   
\begin{equation}
\label{eq_node_simi}
{S_n} = {2^{L + \sum\limits_{l = 1}^L {[\frac{{F({N^l},C_N^l)}}{{F(N_m^l,C_{Nm}^l)}} \times {R_t}]} }}
\end{equation}
where $L$ denotes the max level of Common Nodes, ${\frac{{F({N^l},C_N^l)}}{{F(N_m^l,C_{Nm}^l)}}}$ denotes the \emph{normalized level node similarity factor}.

Here we provide the consideration to introduce the node similarity. Unlike the path similarity measurement used to accurately depict the malicious behavior, the node similarity is considered to facilitate robust detection ability.
Although the path similarity can precisely measure a similar path, it is not flexible enough to capture rough similar patterns, especially the dynamic packets. Generally, dynamic packets also exist in the handshake procedure with automated packets, like the inspection results of the victim machine transferred by the RAT client to the server. These dynamic packets can truncate the CWP for their random DirPiz. Hence, we propose the node similarity as a supplementary to path similarity to handle the dynamic packets problem.

Following the example in previous section, we provide CWP and Common Nodes of former MLTree instances in \autoref{Fig_total_similarity}. Based on these, the similarity score of the example can be calculated, $S_{fp} = 69.13$, $S_{fn} = 80.63$, $S_{tp} = 34.56$, $S_{tn} = 69.12$, when setting $\alpha = 0.3$, and $\beta = 0.7$, $S = 71.65$. 

\subsection{Prediction} \label{sec:prediction}

\begin{figure}
\centerline{\includegraphics[width=1\columnwidth]{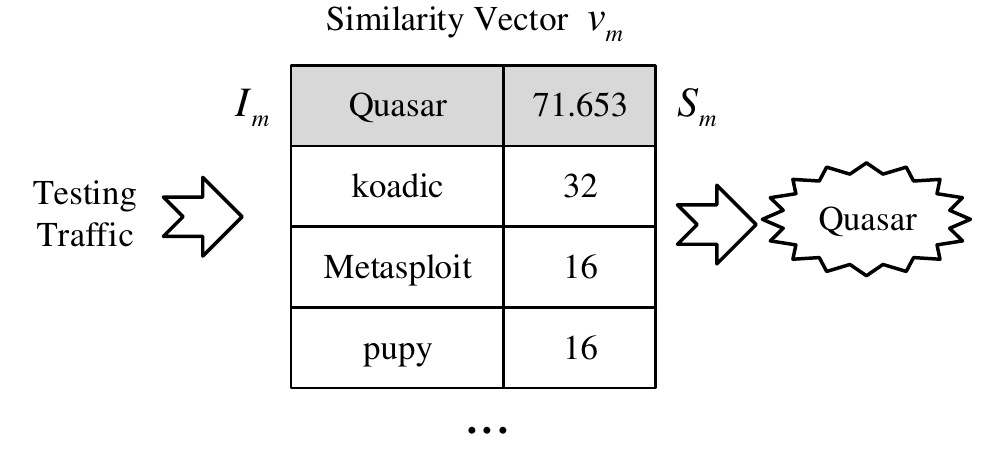}}
\caption{Prediction logic. $\theta$ is set as 32.}
\label{Fig_maxlogic}
\end{figure}
After similarity calculation, we acquire the similarity vector $v_m$ denoting the similarity of the testing instance with each signature. 
Hence, we can further predict if the testing instance is malicious based on the max value $S_m$ of the vector $v_m$. First, the $S_{m}$ is specified. Then, if $S_m$ exceeds a predefined threshold $\theta$, the testing instance is regarded as containing malicious behaviors. Otherwise, the testing instance is regarded as benign. \autoref{Fig_maxlogic} shows the prediction logic with setting $\theta$ as 32.
In addition, in the situation that the testing instance is regarded as malicious, the specific type of malicious behavior can also be predicted based on the index of the malicious value $I_m$.
It is worth to mention that the specific encryption RAT type will only be decided when $S_m$ exceeds the threshold $\theta$. Otherwise, it is meaningless to predict the specific RAT type since the traffic is regarded as benign.

\section{Theoretical Analysis}
To demonstrate the effectiveness of our proposed methods, we provide theoretical analysis in this section. First, since the principle of MBTree is based on the uniqueness of the DirPiz sequence, we demonstrate the extremely low probability that two different applications would produce the same DirPiz sequences. Assume that the DirPiz $X$ at a level are random variables, $P(X)$ represent the probability dense function of $X$, and a DirPiz sequence $S$ is consists of several random variable $S^{m} = [X_0, X_1, ..., X_m]$, where $X_0, X_1, ..., X_m$ are independent. Due to the limit of MTU on the Internet, X can be only chosen from [-1500, 1500] in most cases. 
According to \cite{castro2013probability} and \cite{mccreary2000trends}, $X$ on the Internet generally obey to the $\beta$ distribution. However, after a small amount of frequent DirPiz are removed, $X$ roughly obey to the uniform distribution. That is \[X \sim U( - 1500,1500).\]
Hence, when give two independent DirPiz sequences $S^{m}$ and $S'^{m}$, the probability of their collision in $n$ levels $P(S_n^{m})$ is $C(m, n) * (3*10^{-3})^n$. Suppose that we take $m$ as 10, the corresponding $P(S_n^{10})$ shown in \autoref{tab_probability}. As can be seen from the table, the probability of $S^{10}$ and $S'^{10}$ being completely the same is only $10^{-26}$ in the case of $m=10$, which is extremely low. 
It is worth to mention that, previous studies uses DirPiz snippets at arbitary position rather than corresponding hierarchy to identify application behavior, which increases the collision probability. 

\begin{table}[h]
\centering
\caption{The probability of different $n$ collision in two DirPiz sequences.}
\label{tab_probability}
\begin{tabular}{cc|cc}
\hline
$n$ & $P(S_n^{10})$          & $n$  & $P(S_n^{10})$         \\ \hline
1 & $3\times 10 ^ {-2}$  & 6  & $1\times 10 ^ {-13}$ \\
2 & $4\times 10 ^ {-4}$  & 7  & $2\times 10 ^ {-16}$ \\
3 & $3\times 10 ^ {-6}$  & 8  & $3\times 10 ^ {-19}$ \\
4 & $1\times 10 ^ {-8}$  & 9  & $2\times 10 ^ {-22}$ \\
5 & $6\times 10 ^ {-11}$ & 10 & $6\times 10 ^ {-26}$ \\ \hline
\end{tabular}
\end{table}

Second, based on the collision probability, we discuss the theoretical best threshold interval. Generally, the interval is decide based on number of unique applications $N_A$. When $N_A$ is ensured, the tolerable level of collisions should determined to calculate the max threshold $\theta$. Specifically, $\theta = 2^{L+n}$, where n should satisfy the following conditions \[\left\{ \begin{array}{l}
{N_A} \times P(S_n^m) \le {\rm{1}}\\
{\rm{max(n)}}
\end{array}. \right.\]  For example, when $N_A =100$, the optimal value of $n$ is 1, and the corresponding threshold should be $2^{L+1}$. 

\begin{table*}[]
\centering
\footnotesize
\caption{This table shows the partitioning details in each dataset. The $\bullet$ and $\star$ denotes the OSER traffic generated from Debian and CentOS respectively; $\triangleright$ denotes the WT traffic; $\otimes$ denotes the CTU-13 traffic; $\square$ and $\circ$ denote different applications traffic from the combined set of ISCX VPN2016 and USTC-TFC2016. }
\label{tab_data_partition}
\begin{tabular}{c|c|c|c|c|c|c|c|c|c}
\hline
          & \multicolumn{3}{c|}{Dataset I}   & \multicolumn{3}{c|}{Dataset II}                        & \multicolumn{3}{c}{Dataset III}                    \\ \hline
          & Train     & Validation & Test    & Train            & Validation       & Test             & Train           & Validation      & Test            \\ \hline
Malicious & $\bullet$ & $\bullet$  & $\star$ & $\triangleright$ & $\triangleright$ & $\triangleright$ & $\otimes$ & $\otimes$ & $\otimes$ \\ \hline
Benign    & $\square$ & $\square$  & $\circ$ & $\square$        & $\square$        & $\circ$          & $\square$       & $\square$       & $\circ$         \\ \hline
\end{tabular}
\end{table*}
\section{Evaluation Framework}

\subsection{Evaluation Data}
In our experiment, three malicious parts and two benign parts of traffic are used for evaluation. The first malicious part is the open-source RATs traffic collected by ourselves, the second part is the wild Trojan traffic selected from the Stratosphere project \cite{Stratosphere}, and the last part is a public open-source dataset CTU-13 \cite{garcia2014empirical}. While the two parts benign applications traffic used are from two open datasets, ISCX VPN2016 \cite{draper2016characterization} and USTC-TFC2016 \cite{wang2017malware}.
Each part is described below.

\begin{figure}
\centerline{\includegraphics[width=0.7\columnwidth]{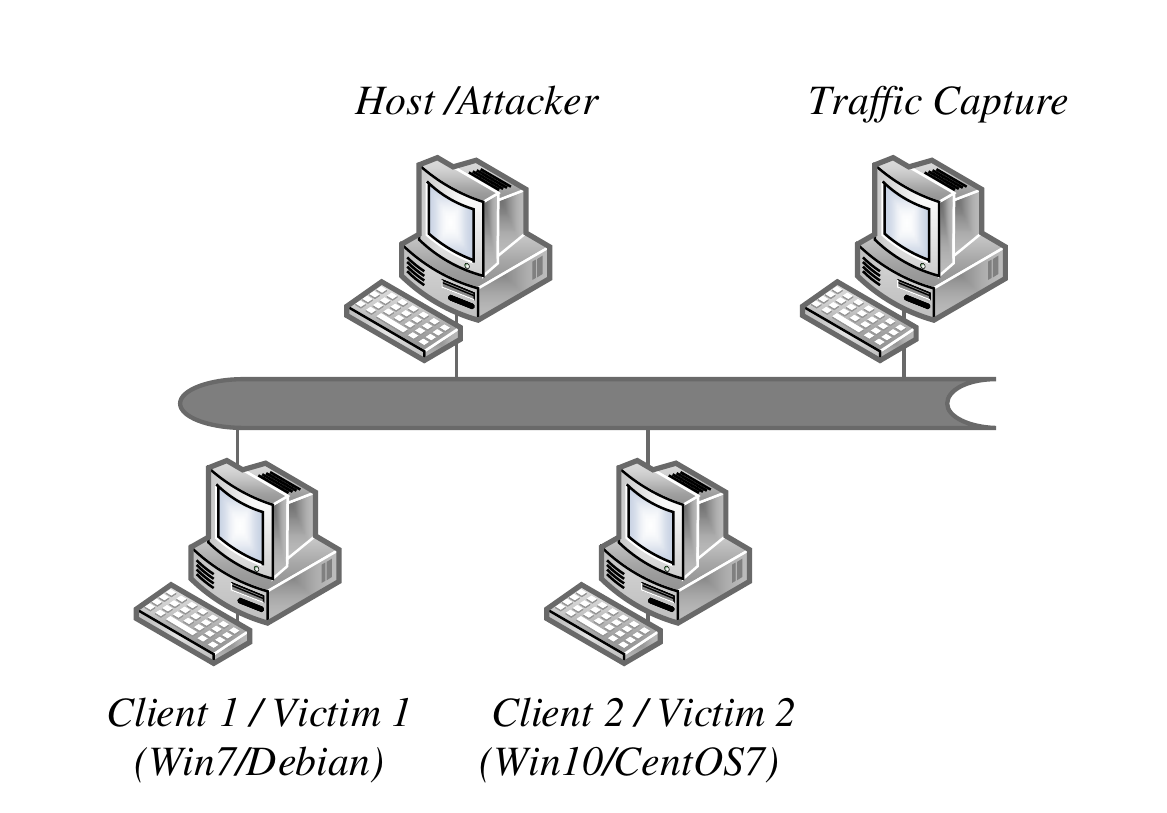}}
\caption{Traffic collection mechanism.}
\label{Fig_rat_collect}
\end{figure}

\textbf{Open Source Encryption RAT (OSER);}  In order to hide the real identity, adopting customized OSER for attack is popular in recent years \cite{fuelRAT, wueest2016increased}. Thus, the open-version RAT traffic is studied in this paper. Based on popularity, stability, and maintenance on Github, 7 OSERs are selected to generate this part of the traffic.
Specifically, in the traffic generation procedure, to evaluate whether a RAT follows the same procedure for communication in different environments, we collect the traffic of two hosts, which install different systems but are infected by the same sample.
The traffic generation mechanism is shown in \autoref{Fig_rat_collect}. Besides, to simulate the practice usage of samples, 5 randomly chosen commands are executed on the comprised machine for each malicious session.
The details of the collected traffic are shown in \autoref{tab_rats} in Appendix.

\textbf{Wild Trojan (WT);} Apart from the OSER traffic, wild trojans (from 2015 to 2018) are also selected from \cite{Stratosphere}. Compared to OSER traffic, the WT traffic contains more number sessions. However, since the communications between the victim and Trojan C\&C are not controlled as detailed as OSER, it may also contain noise traffic generated by the machine automatically. Nevertheless, this part of the traffic is also a fair test to evaluate the WT detection ability of MBTree. More details of this part of the traffic are shown in \autoref{tab_trojans} in Appendix.

\textbf{CTU-13;} The public open-source dataset is widely used in malware traffic detection research \cite{wang2016botnet, ring2019survey}. It contains malicious traffic from 7 different types of Trojans. Apart from C\&C traffic, it also collects attack traffic in other procedures, like distributed denial of service (DDoS) traffic and spam traffic. Hence, this traffic is cleaned to only reserve C\&C traffic according to the dataset description. Due to the lack of valid traces for 5-fold validation, two types of Trojan (Rbot, NSIS.ay) are not used in our experiment. More details of cleaned traffic are shown in \autoref{tab_ctu13} in Appendix. 

\textbf{ISCX VPN 2016;} In recent studies, the ISCX VPN2016 is widely used for encrypted traffic classification. In this paper, we also adopt this set as a part of benign traffic to evaluate the false alarm levels.
Totally, the set organizes the 27G raw traffic generated from 17 typical applications in 150 pcap or pcapng files. Compared to former malicious parts, this benign application set is larger.

\textbf{USTC-TFC 2016;} Apart from ISCX VPN2016, we also use the benign part of USTC-TFC2016 as another part of benign traffic, because it contains different applications traffic from the ISCX VPN2016. Actually, this part of traffic consists of both benign and malicious traffic. Since the traffic contained in the malicious part of USTC-TFC2016 is covered by former datasets, they are not taken as malicious parts. Thus, only the benign part of the set is used for evaluation. Totally, the benign part of the USTC-TFC2016 organizes the 3.71GB traffic generated from 10 applications in 14 pcap files.

\subsection{Data Organization}
We organize the five parts of the collected traffic into three datasets for evaluation. Each dataset consists of malicious traffic and benign traffic. The OSER is used as malicious traffic in dataset I, the WT is used in dataset II, and CTU-13 is used in dataset III. The benign traffic shared by two datasets is the combination of ISCX VPN 2016 and USTC-TFC 2016. Total $N_A$ of these two datasets is around 100.

For persuasive evaluation, we use the 5-fold cross-evaluation strategy to acquire a stable performance. Moreover, in each fold, the dataset is divided into three parts, train, validation, and test. Basically, the three parts in a fold are divided following the ratio of 0.49:0.21:0.3. It is noteworthy that since OSER is generated from two machines, we use the traffic generated from one machine as the train and validation sets respectively, and from the other machine as the test set. Besides, the benign applications that appeared in the train set do not appear in the test set. This partitioning strategy aims at simulating the situation that unknown applications that emerged in the test environment, which scenario is common in reality. The details of partition are shown in \autoref{tab_data_partition}.

\begin{table*}
\centering
\caption{Experiment results. Highlighted values: \textbf{overall best method}, second best method ($\dagger$), third best method ($\ddagger$).}
\label{tab_re}
\begin{tabular}{c|c|cccc|cccc} 
\hline
Dataset & Methods & \multicolumn{4}{c|}{Validation} & \multicolumn{4}{c}{Test} \\ 
\hline
 &  & FPR & FNR & Acc & F1 & FPR & FNR & Acc & F1 \\ 
\hline
\multirow{7}{*}{I} & \underline{MBTree}  & \underline{0.21±0.21} $\dagger$  & \underline{0.23±0.35}  & \underline{99.82±0.23} $\ddagger$  & \underline{99.27±0.12} $\dagger$  & \textbf{\underline{0.1±0.09}}  & \underline{1.41±1.55} $\dagger$  & \textbf{\underline{99.59±0.25}}  & \textbf{\underline{99.42±0.15}}  \\
 & CART & 2.49±5.05 & 0.03±0.07 $\dagger$  & 99.56±2.68 & 95.71±11.54 & 26.77±6.99 $\ddagger$  & 1.73±2.1 $\ddagger$  & 74.2±31.43 $\ddagger$  & 66.17±8.43 $\ddagger$  \\
 & RF & 24.79±19.45 & 0.08±0.07 $\ddagger$  & 99.36±0.93 & 74.59±20.34 & 33.06±16.54 & 3.2±1.28 & 57.44±21.63 & 57.72±21.94 \\
 & GBDT-CIC & 49.25±22.02 & 1.59±3.11 & 79.2±44.25 & 47.37±25.22 & 67.16±13.82 & 3.85±2.2 & 54.02±21.56 & 25.39±11.93 \\
 & CNN & \textbf{0.03±0}  & \textbf{0.01±0}  & \textbf{99.98±0}  & \textbf{99.95±0.06}  & 70.6±1.14 & 2.4±0.43 & 65.93±7.53 & 30.77±0.79 \\
 & SAE & 1.37±2.98 $\ddagger$  & \textbf{0.01±0}  & 99.97±0.03 $\dagger$  & 98.57±3.16 $\ddagger$  & 67.85±3.08 & 2.32±0.55 & 68.94±9.44 & 33.65±3.86 \\
 & DirPiz-Seq & 18.49±2.29 & 1.45±0.66 & 80.28±9.77 & 83.14±1.95 & 18.1±2.42 $\dagger$  & \textbf{0.99±0.2}  & 84.52±3.09 $\dagger$  & 80.66±1.83 $\dagger$  \\ 
\hline
\multirow{7}{*}{II} & \underline{MBTree}  & \underline{\textbf{0.57±1.14}}  & \underline{\textbf{0.03±0.05}}  & \textbf{\underline{99.96±0.06}}  & \underline{92.43±1.33} $\ddagger$  & \underline{\textbf{0.62±1.25}}  & \underline{0.84±0.91} $\ddagger$  & \underline{\textbf{99.08±0.94}}  & \underline{\textbf{87.71±1.71}}  \\
 & CART & 6.05±2.18 $\ddagger$  & 0.81±0.26 & 95.72±1.05 & 93.14±1.51 $\dagger$  & 52.66±13.2 & 5.43±2.23 & 68.06±17.96 & 44.17±16.88 \\
 & RF & 10.32±13.01 & 1.4±1.5 & 91.62±10.19 & 88.01±17.97 & 38.67±9.98 & 11.77±12.91 & 67.68±13.00 & 51.93±10.99 \\
 & GBDT-CIC & 3.56±1.63 $\dagger$  & 0.57±0.51 $\dagger$  & 96.94±0.74 $\dagger$  & \textbf{96.33±1.32}  & 38.78±11.82 & 6.49±3.21 & 65.00±15.69 & 52.54±9.44 \\
 & CNN & 15.67±1.99 & 0.66±0.33 $\ddagger$  & 95.75±2.16 $\dagger$  & 83.43±2.28 & 19.79±7.08 $\ddagger$  & \textbf{0.58±0.25}  & 95.56±1.88 $\dagger$  & 74.4±7.20 $\ddagger$  \\
 & SAE & 15.74±2.6 & 0.78±0.51 & 95.16±3.28 & 83.05±3.14 & 19.13±6.78 $\dagger$  & 0.59±0.41 $\dagger$  & 95.46±3.1 $\ddagger$  & 75.71±9.35 $\dagger$  \\
 & DirPiz-Seq & 40.27±4.77 & 7.55±1.69 & 58.20±9.39 & 56.33±3.70 & 41.96±5.65 & 7.59±0.41 & 52.84±2.15 & 52.00±2.28 \\ 
\hline
\multirow{7}{*}{III} & \underline{MBTree}  & \textbf{\underline{0.0±0.0}}  & \textbf{\underline{0.0±0.0}}  & \underline{99.99±0.01} $\dagger$  & \underline{91.11±0.0}  & \textbf{\underline{0.0±0.0}}  & \underline{0.12±0.13} $\ddagger$  & \underline{\textbf{99.88±0.13}}  & \underline{\textbf{86.66±6.67}}  \\
 & CART & 0.17±0.24 $\dagger$  & 0.22±0.35 $\dagger$  & 99.61±0.27 & 93.59±4.11 & 0.03±0.05 $\dagger$  & 12.29±14.53 & 88.95±13.39 $\dagger$  & 79.58±11.73 $\dagger$  \\
 & RF & \textbf{0.0±0.0}  & \textbf{0.0±0.0}  & 99.92±0.06 & 97.6±0.67 $\ddagger$  & \textbf{0.0±0.0}  & 79.42±13.2 & 21.65±12.9 & 65.4±8.2 \\
 & GBDT-CIC & \textbf{0.0±0.0}  & \textbf{0.0±0.0}  & 99.93±0.05 $\ddagger$  & 98.35±1.1 $\dagger$  & 0.03±0.07 $\ddagger$  & 75.82±20.42 & 25.17±20.07 & 66.49±9.1 \\
 & CNN & 1.36±2.71 $\ddagger$  & \textbf{0.0±0.0}  & 99.2±1.5 & 93.5±7.97 & 82.84±15.09 & 0.01±0.02 $\dagger$  & 81.12±15.04 $\ddagger$  & 70.95±8.02 $\ddagger$  \\
 & SAE & \textbf{0.0±0.0}  & \textbf{0.0±0.0}  & \textbf{100.00±0.0}  & \textbf{100.00±0.0}  & 87.96±8.12 & \textbf{0.0±0.0}  & 72.11±21.1 & 64.46±15.05 \\
 & DirPiz-Seq & 20.85±2.27 & 21.6±38.72 $\ddagger$  & 66.69±17.73 & 52.63±8.78 & 14.92±0.31 & 49.0±31.39 & 61.32±5.18 & 54.04±5.8 \\
\hline
\end{tabular}
\end{table*}

\begin{figure*}[h]
\centering
\subfigure[Confusion matrix on Dataset I. ]{
\label{Fig_cnf_rats}
\includegraphics[width=0.25\linewidth]{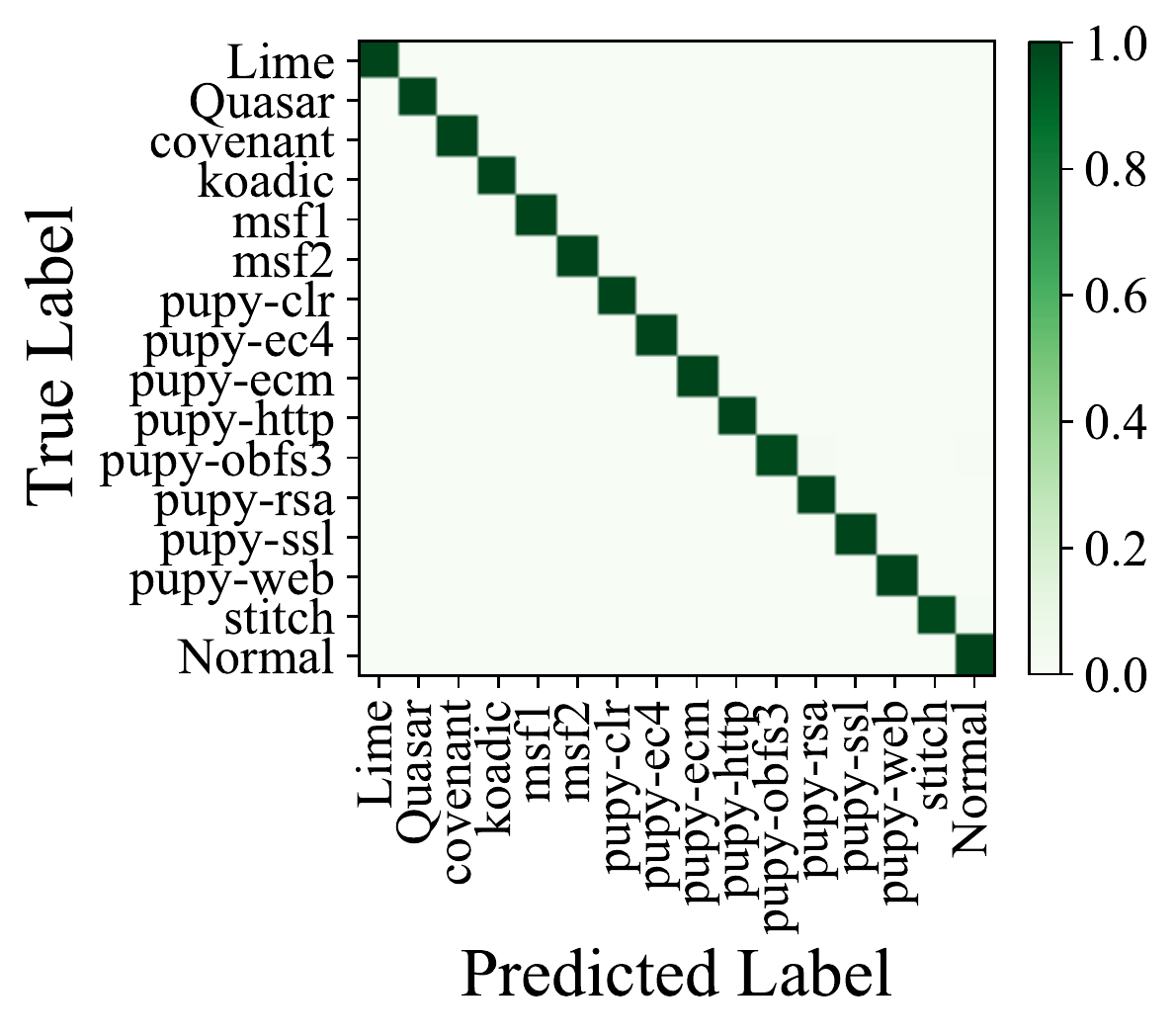}
}
\subfigure[Confusion matrix on Dataset II.]{
\label{Fig_cnf_malware}
\includegraphics[width=0.27\linewidth]{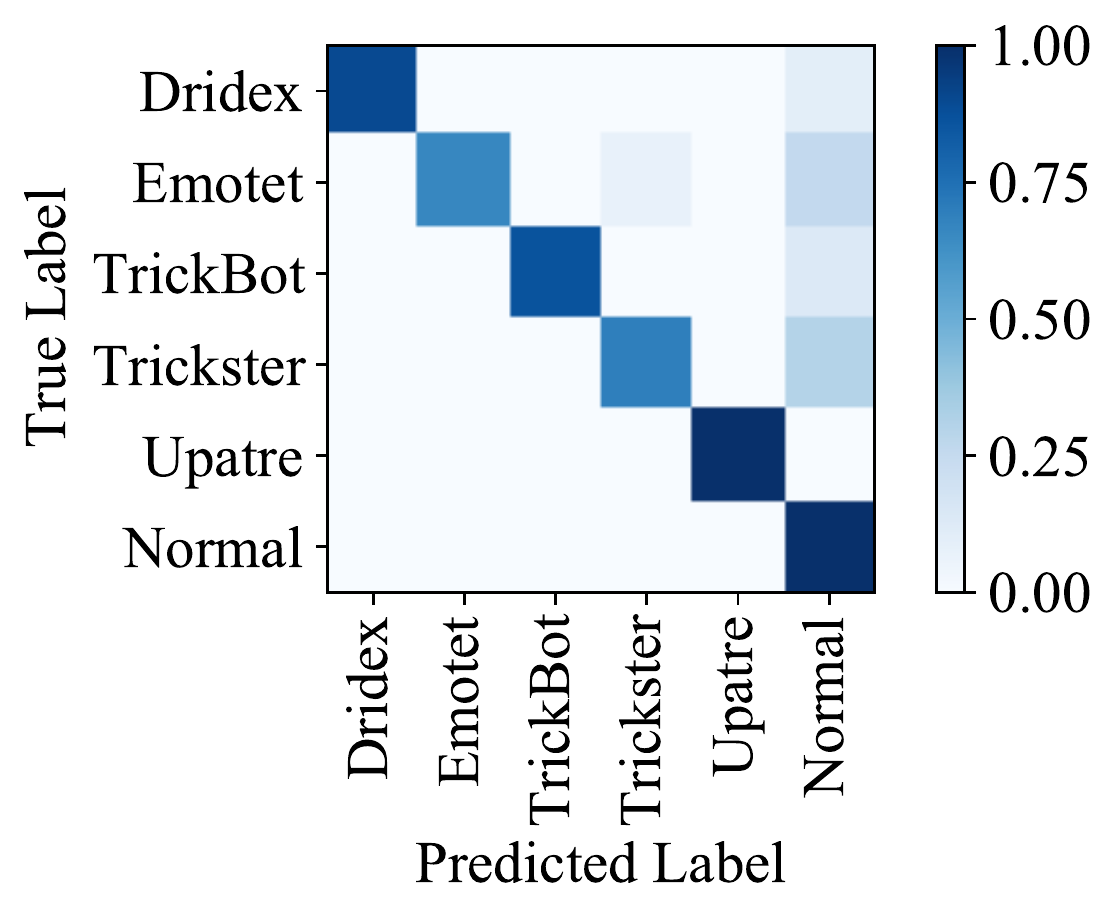}
}
\subfigure[Confusion matrix on Dataset III. ]{
\label{Fig_cnf_ctu}
\includegraphics[width=0.27\linewidth]{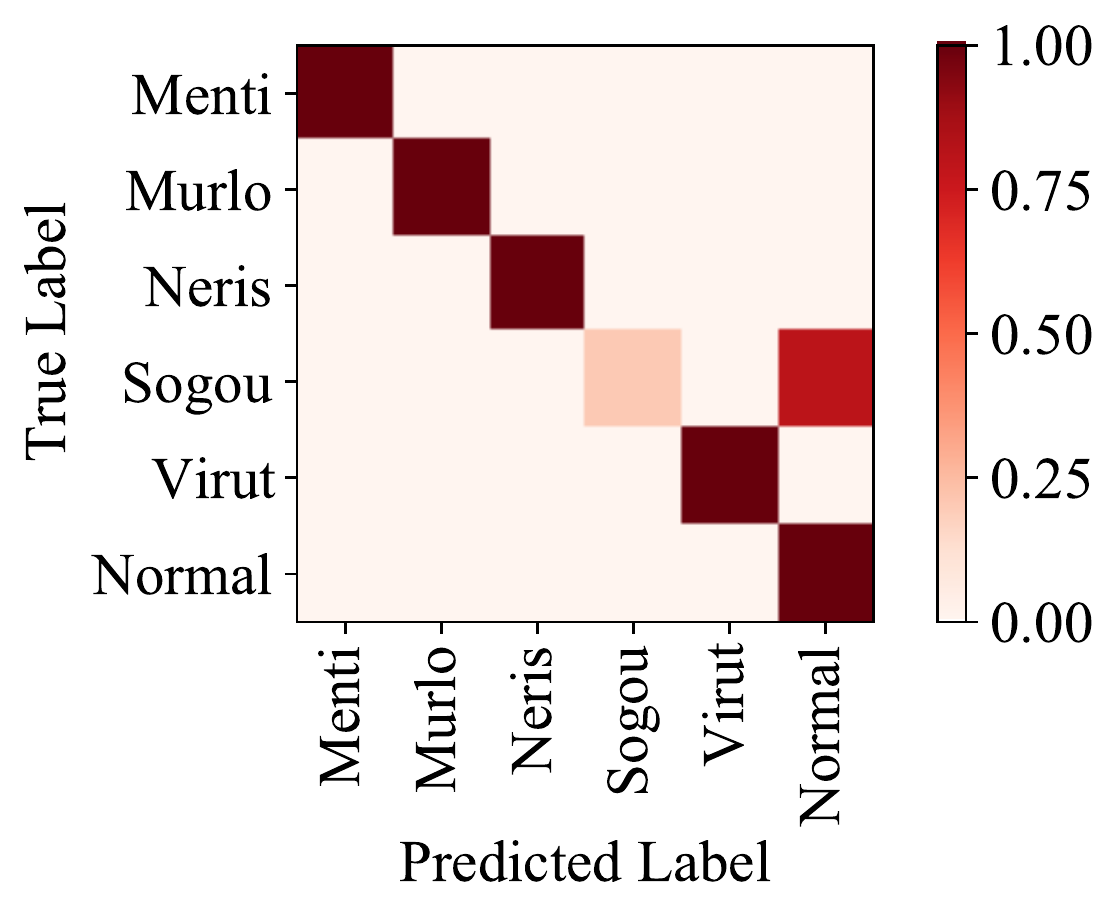}
}
\caption{Confusion matrices of MBTree test predictions on different datasets. } 
\label{Fig_cfx}
\end{figure*}

\subsection{Baselines}
In correspondence to the related studies in \autoref{sec:relatedwork}, six baselines are covered in this paper as comparisons.
The first baseline is a machine learning state-of-the-art \cite{stergiopoulos2018automatic} using side-channel features and CART represented as \emph{CART}.
The second baseline is another machine learning-based state-of-the-art \cite{gezer2019flow} using different features from \cite{stergiopoulos2018automatic} with random forest for Trickbot Trojan detection as \emph{RF}.
The third baseline is the extended features generated by CICFlowmeter \cite{CICFlowmeter} with GradientBoosting implemented by ourselves as \emph{GBDT-CIC}.
The fourth baseline is a deep learning method using one dimensional CNN from \cite{lotfollahi2020deep} as \emph{CNN}.
The fifth baseline is another deep learning method using a stacked autoencoder mechanism as \emph{SAE} \cite{lotfollahi2020deep}.
Besides, we also implement the flow-level similar matching method using the cosine distance and threshold of 0.99 represented as \emph{DirPiz-Seq} to illustrate the advantage of the host-level signatures.
Moreover, we try to compare our method with ping-pong \cite{trimananda2019pingpong}, which is a signature-based state-of-the-art for encrypted IOT traffic event identification. However, we find that ping-pong can hardly extract packet-level signatures of malicious traffic. Because the conversation pairs of packet-level signatures are too diverse to be clustered by the DBSCAN algorithm even though we try to tune corresponding parameters.

\subsection{Tasks \& Metrics}
Based on the aforementioned data, we propose the detection task for experiments. The goal of the task is to distinguish whether the traffic is benign or malicious. Specifically, the malicious traffic is labeled as the specific type when is predicted. While for benign traffic, they are all regarded equally as general benign without further classifying the specific application type.
In addition, we also record the prediction time of each method.
As evaluation, four well-known metrics are adopted: \emph{False Positive Ratio (also known as False Alarms Ration, FPR)}, \emph{False Negative Ratio (FNR)} \cite{shah2018performance}, \emph{Accuracy (Acc)}, and \emph{Macro F1 (F1)}. It should be noted that when calculating FPR and FNR, the multi-class labels are masked to binary labels as only simple malicious or benign. While calculating Acc and F1, the multi-class labels are directly taken into consideration. We believe this mechanism can measure the performance of different models more comprehensively.

\section{Experiments}
In this section, we provide a thorough evaluation of MBTree from five perspectives based on the aforementioned dataset. First, we perform the evaluation on the detection task to show the effectiveness of the proposed MBTree by comparing it to several machine learning-based state-of-the-arts. Second, we provide a case study to clearly show the significance of MBTree. Third, we compare the efficiency of each method. Fourth, we conduct experiments to analyze the influence of the hyperparameters and their corresponding optimal choice. Further, we show the analysis of the generated signatures to reveal the network behavior differences among different samples.
Except for the fourth experiment, the hyperparameters of MBTree are set preliminary as \emph{max level} $L$ of 10, \emph{path similarity ratio} $\alpha$ of 0.3, \emph{head signature ratio} $\beta$ of 0.7, and \emph{threshold} $\theta$ as 2048.

\subsection{Malicious Detection}
In this section, we focus on presenting the overall effectiveness of MBTree. \autoref{tab_re} shows the performance of different approaches on each dataset. And \autoref{Fig_cfx} shows the classification results of MBTree on the test part of different datasets.
First, as a high-level performance comparison, although MBTree is not achieving the best performance on the validation set, it outperforms all the considered elements of contrast in the test set in most metrics. This indicates the robust ability of our proposed signature-based approach. Besides, considering differences in performance on validation and test set, it can be concluded that the signature-based methods perform more stable on different sets, including both MBTree and DirPiz-Seq. From another view, although the statistical distribution of malicious part is consistent in validation and test of dataset II and III, the statistical-based methods also perform quite differently. This illustrates that these statistical-based methods are more susceptible to the change of the environment. Hence, the robust detection ability of signature-based methods through different environments is demonstrated.

Second, It should be noted that the F1 of MBTree on dataset II and III is not reaching the level that on the dataset I. According to \autoref{Fig_cnf_malware} and \autoref{Fig_cnf_ctu}, it can be observed that most of the misclassified instances are false negatives. With a detailed analysis of these instances, we found that they consist of shorter communication sequences, which means there are only one or two valid payloads exchanged between the client and C\&C. Thus, without sufficient evidence, MBTree tends to classify them as benign. An interesting phenomenon is that other metrics are consistent in all datasets, which indicates that though there exist misclassified instances, the number of them is still at a low level. This can be attributed to that the host-level detection reduces the number of instances. Hence, even only a few misclassified instances will lead to a significant change in F1.

Third, it can be noticed that the flow-level DirPiz-Seq performs worse than MBTree. This phenomenon indicates that only utilizing the flow-level communication sequences as signatures is not precise enough. Integrating flow fingerprints to host signatures can improve the performance dramatically.

\begin{figure*}[h]
\centering
\subfigure[Path Score Ratio $\alpha$]{
\label{Fig_alpha}
\includegraphics[width=0.23\linewidth]{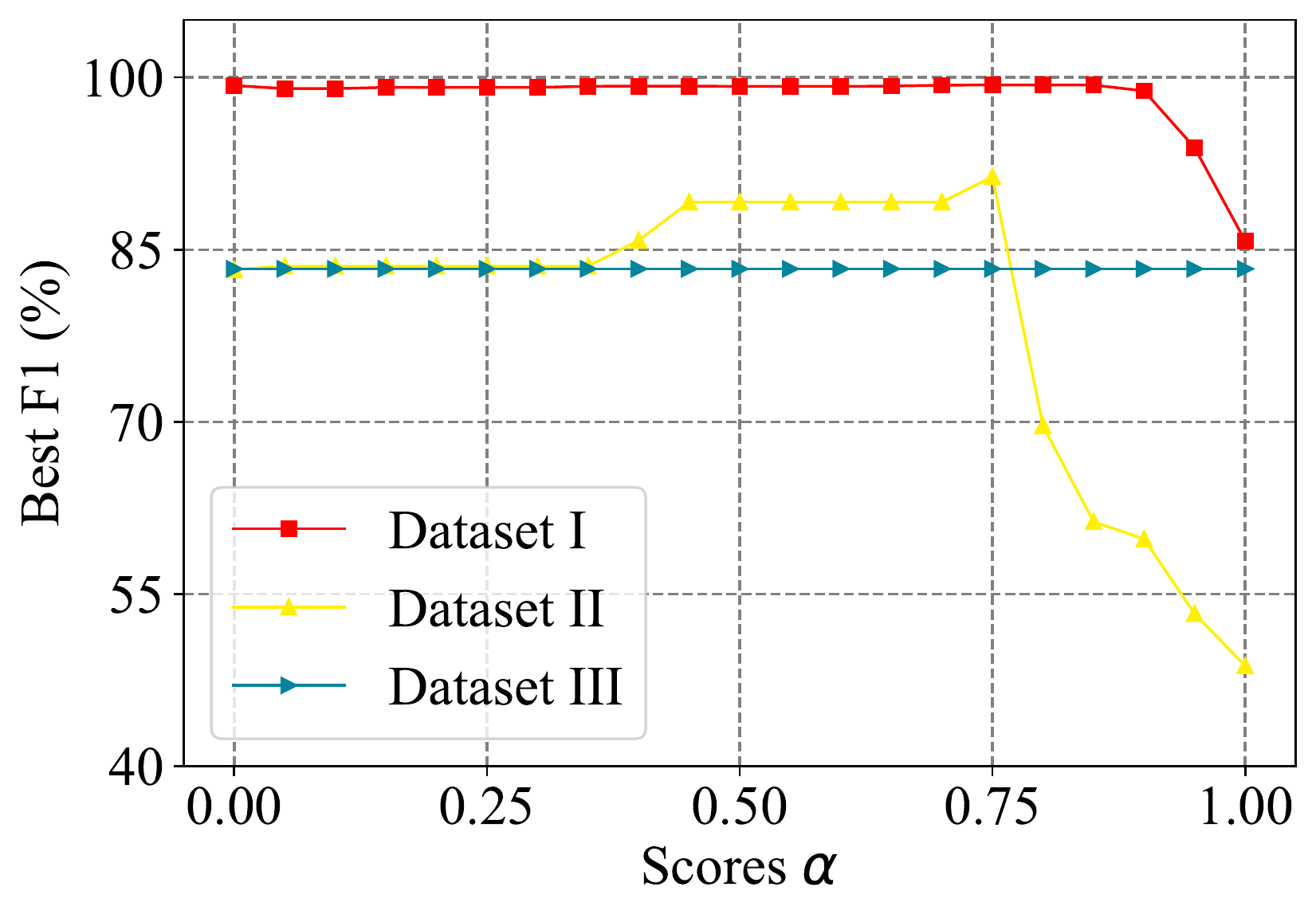}
}
\subfigure[Head Score Ratio $\beta$ ]{
\label{Fig_beta}
\includegraphics[width=0.23\linewidth]{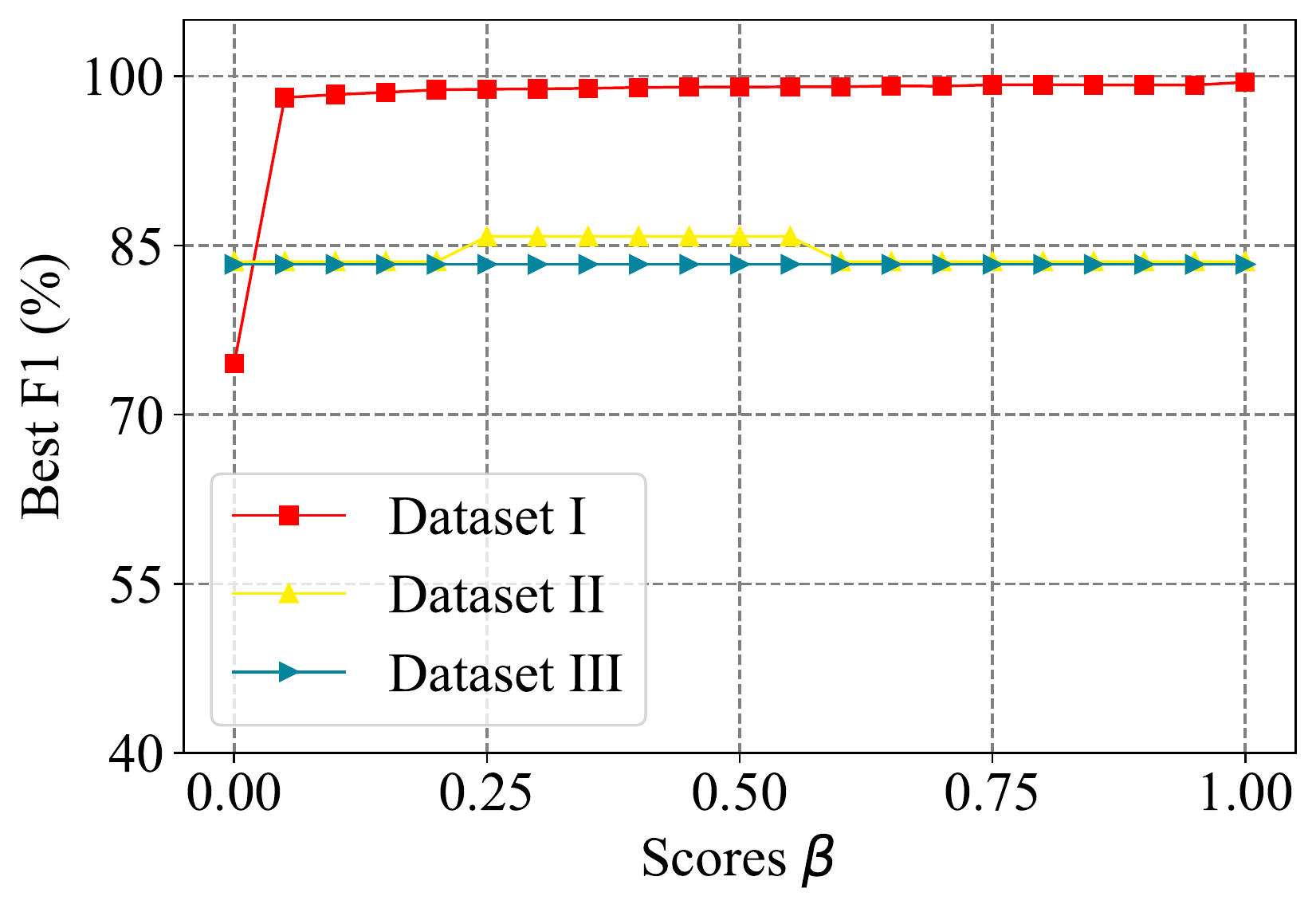}
}
\subfigure[Max Level $L$ ]{
\label{Fig_maxlevel}
\includegraphics[width=0.33\linewidth]{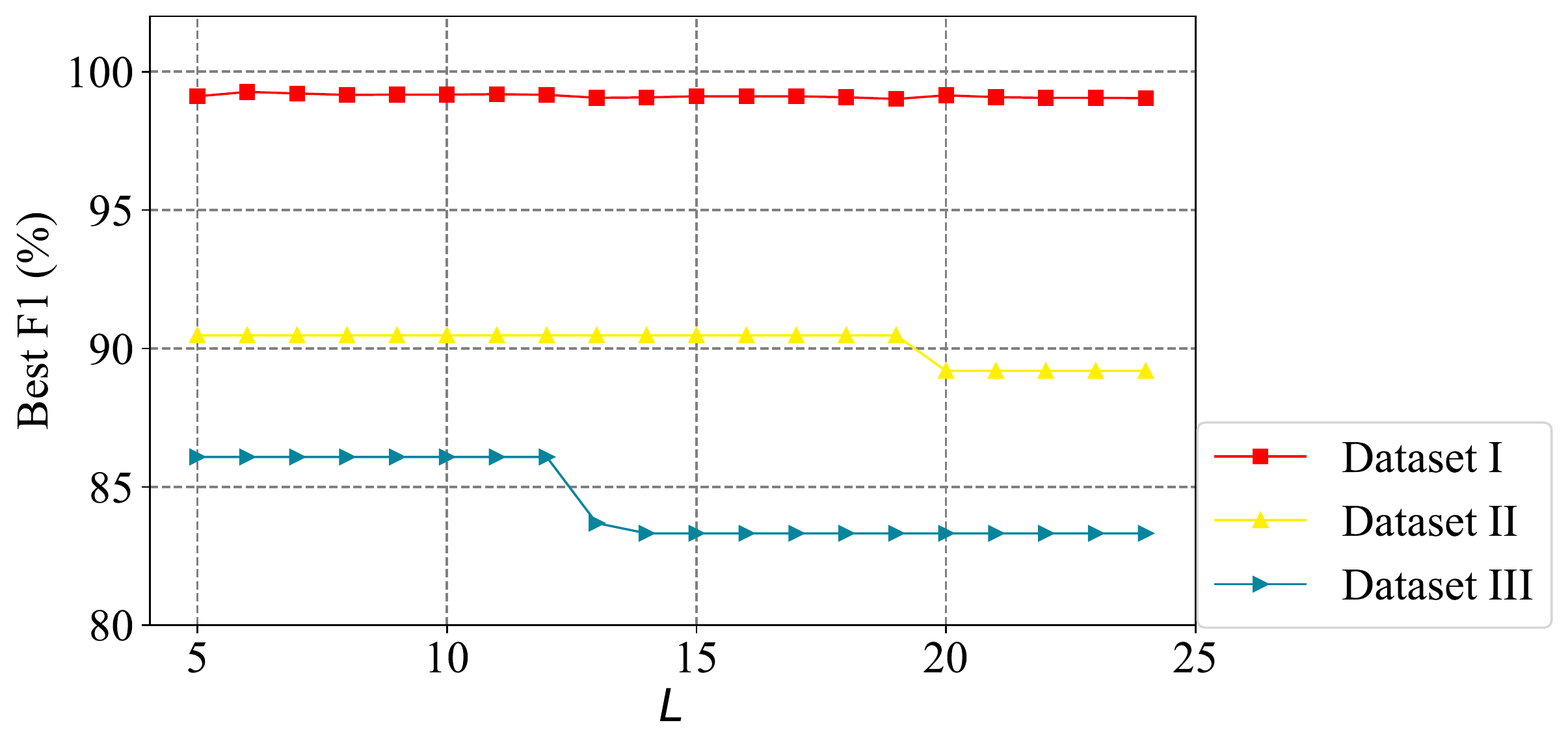}
}
\caption{Different parameter tuning results.} 
\label{Fig_tunning}
\end{figure*}
\begin{figure*}[ht!]
\centering
\subfigure[Max Level $L$ = 5]{
\label{Fig_thresholds_5}
\includegraphics[width=0.2\linewidth]{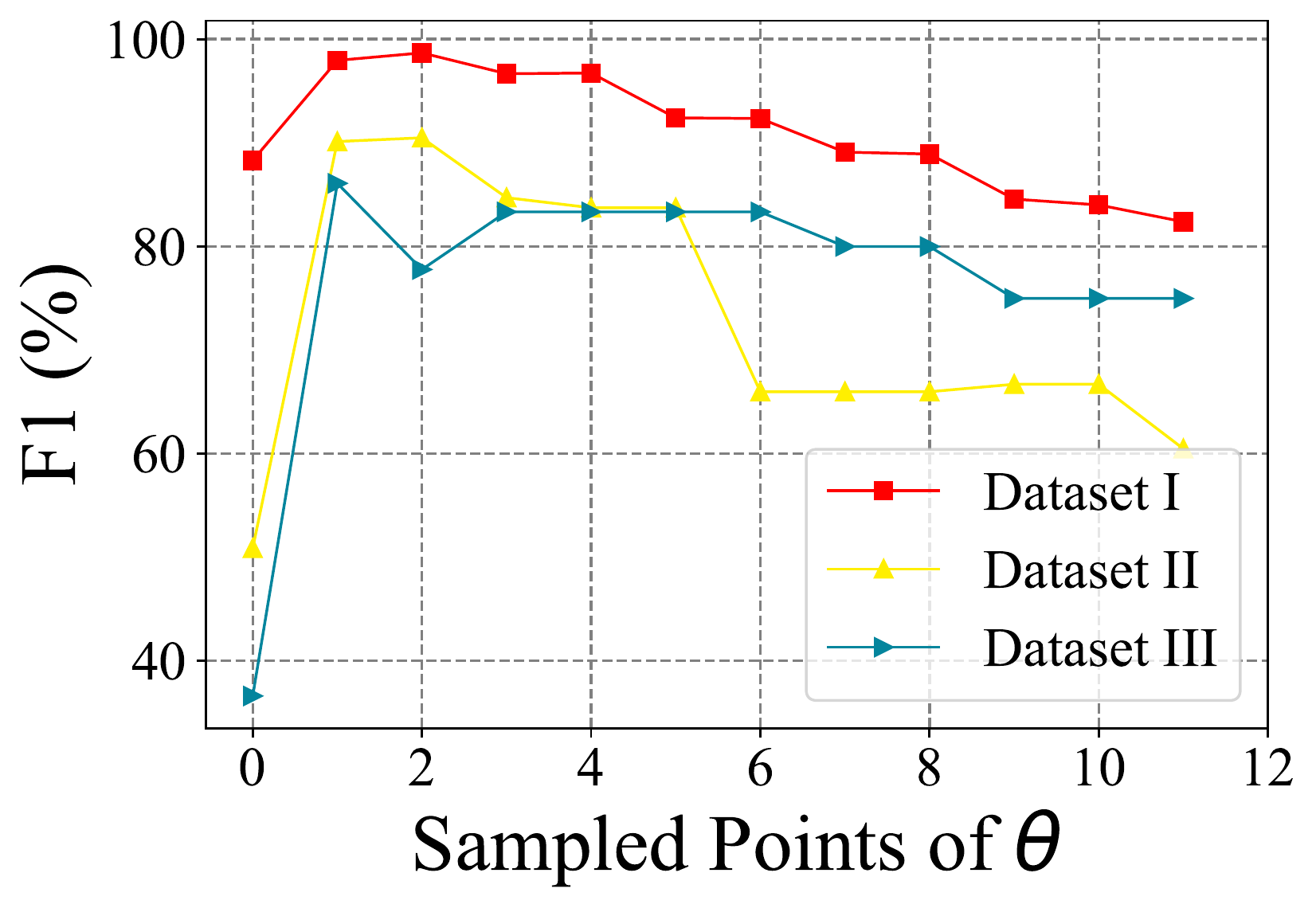}
}
\subfigure[Max Level $L$ = 10 ]{
\label{Fig_thresholds_10}
\includegraphics[width=0.2\linewidth]{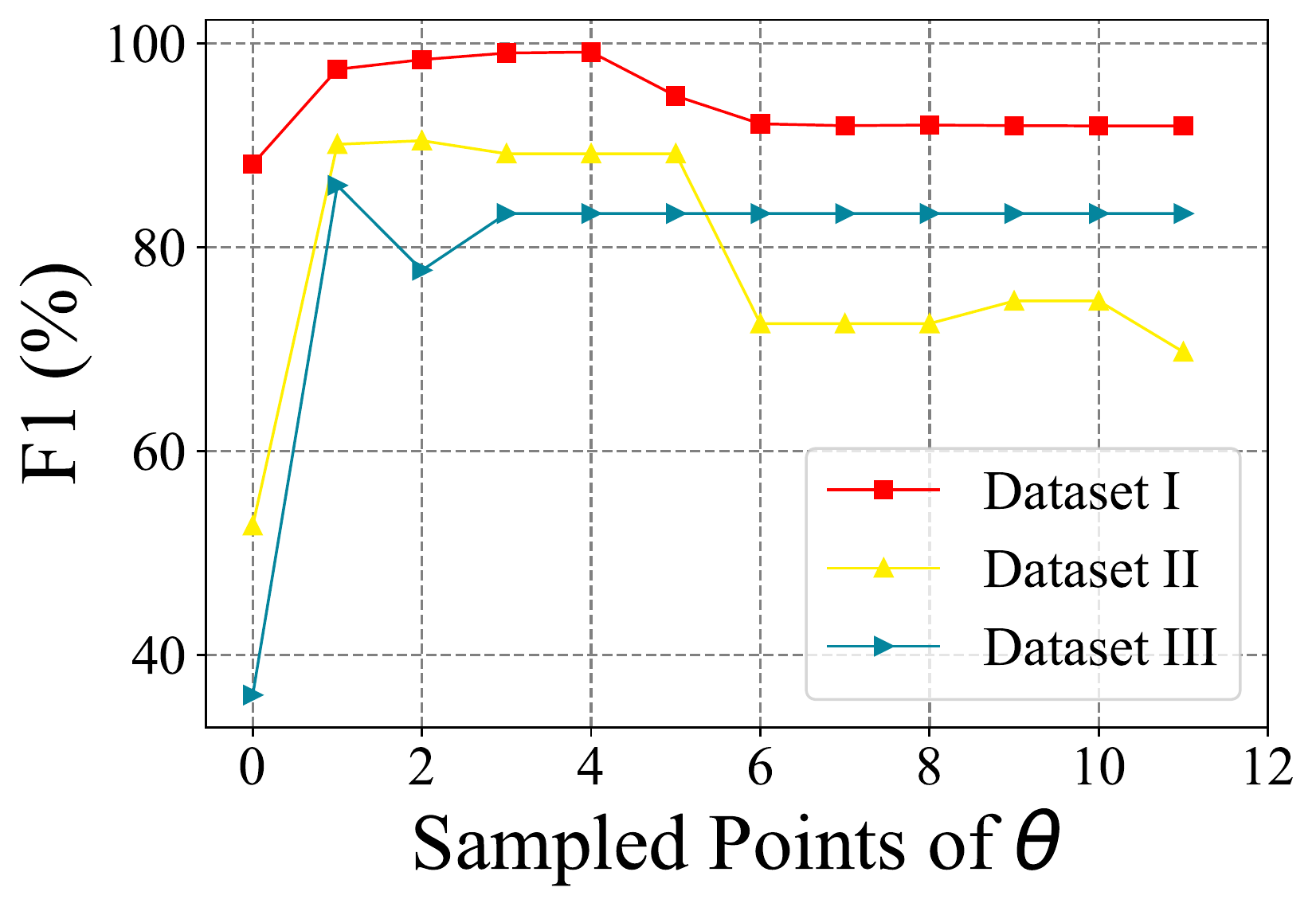}
}
\subfigure[Max Level $L$ = 15 ]{
\label{Fig_thresholds_15}
\includegraphics[width=0.2\linewidth]{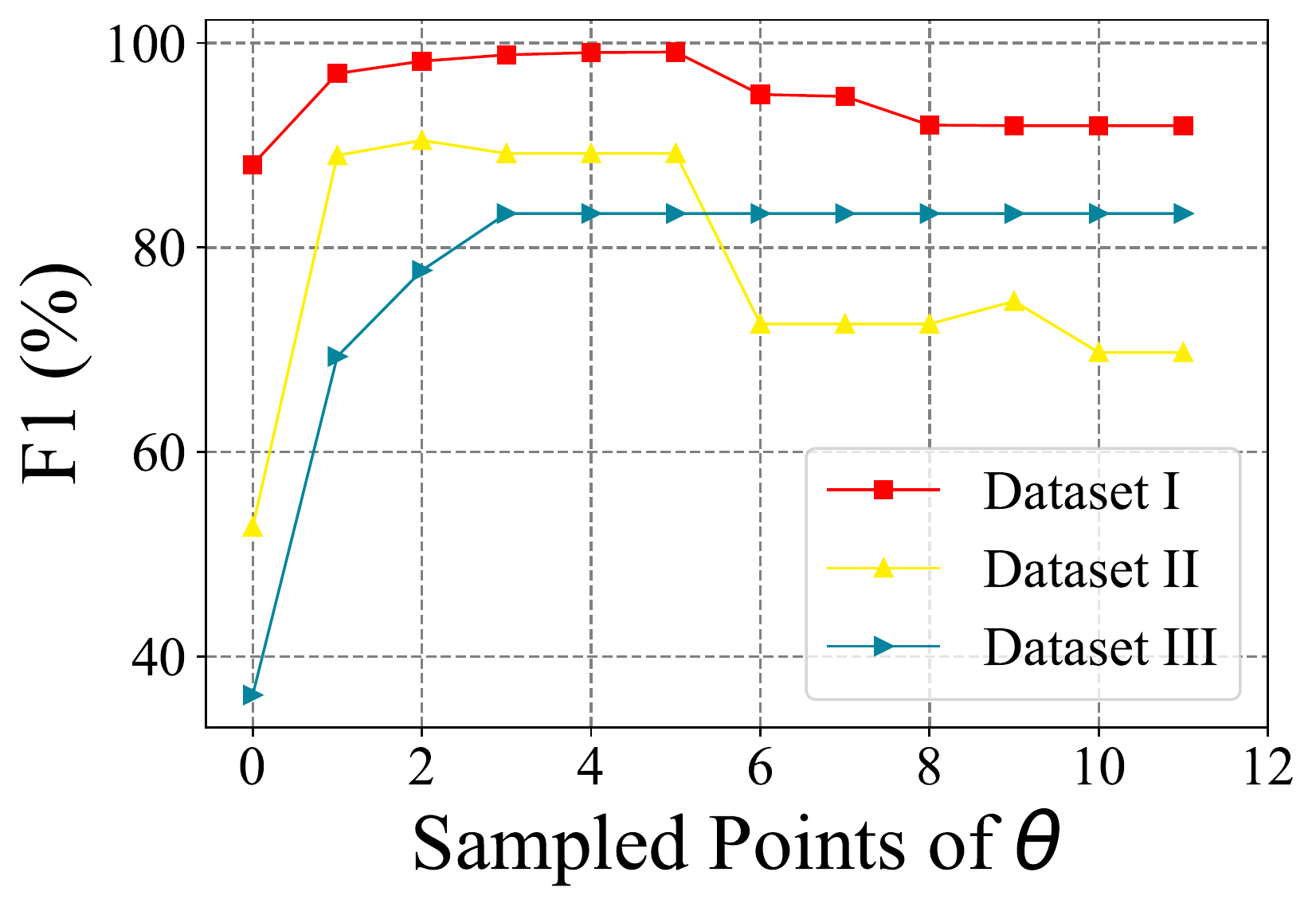}
}
\subfigure[Max Level $L$ = 20]{
\label{Fig_thresholds_20}
\includegraphics[width=0.2\linewidth]{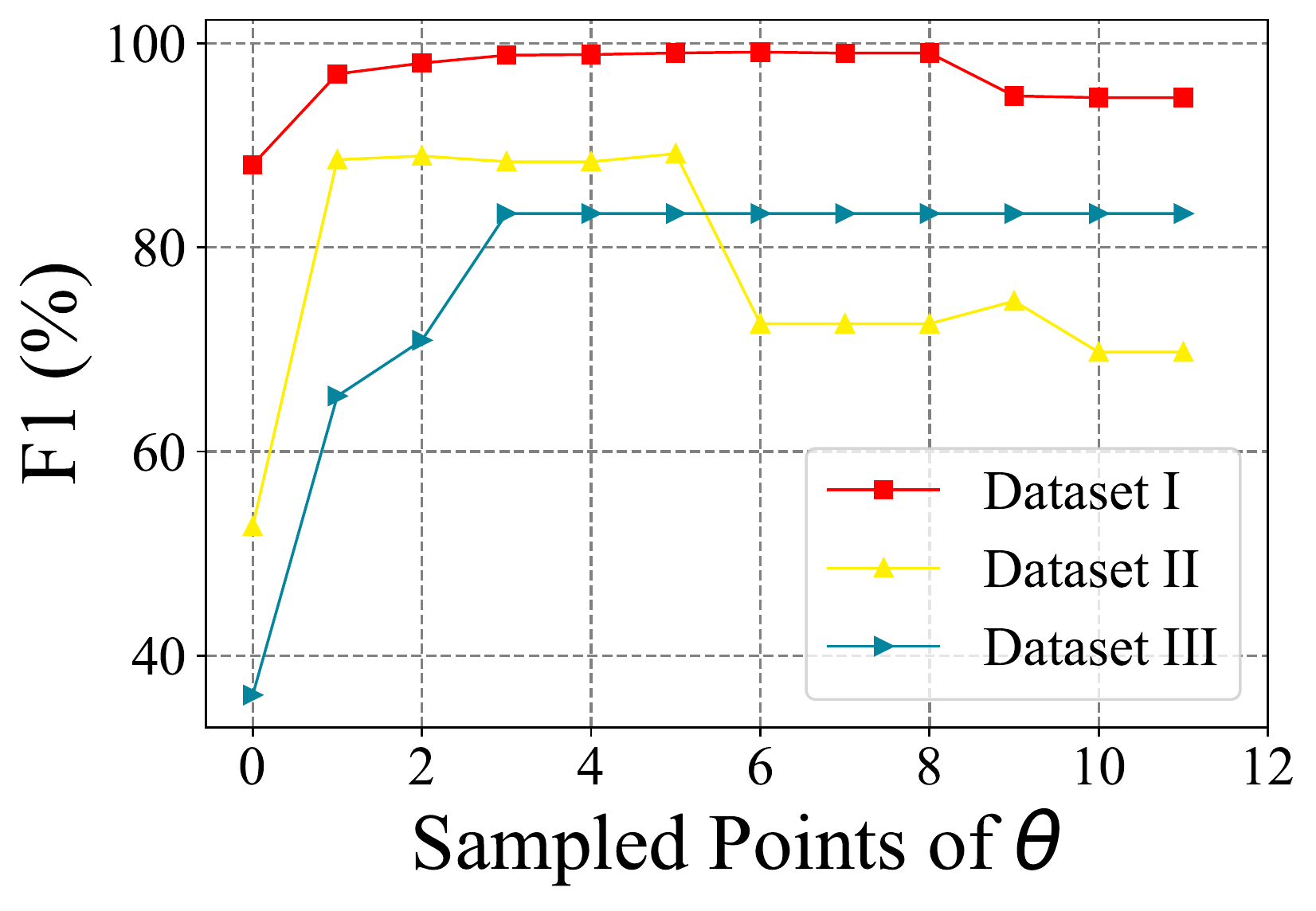}
}
\caption{The tuning results of threshold $\theta$ with different Max Level $L$. For each $L$, 10 points of $\theta$ are sampled between $[2^{L}, 2^{L+2}]$}.
\label{Fig_tunning_threshold}
\end{figure*}
\begin{figure*}[ht!]
\centering
\subfigure[Max Level $L$ = 5]{
\label{Fig_det_5}
\includegraphics[width=0.2\linewidth]{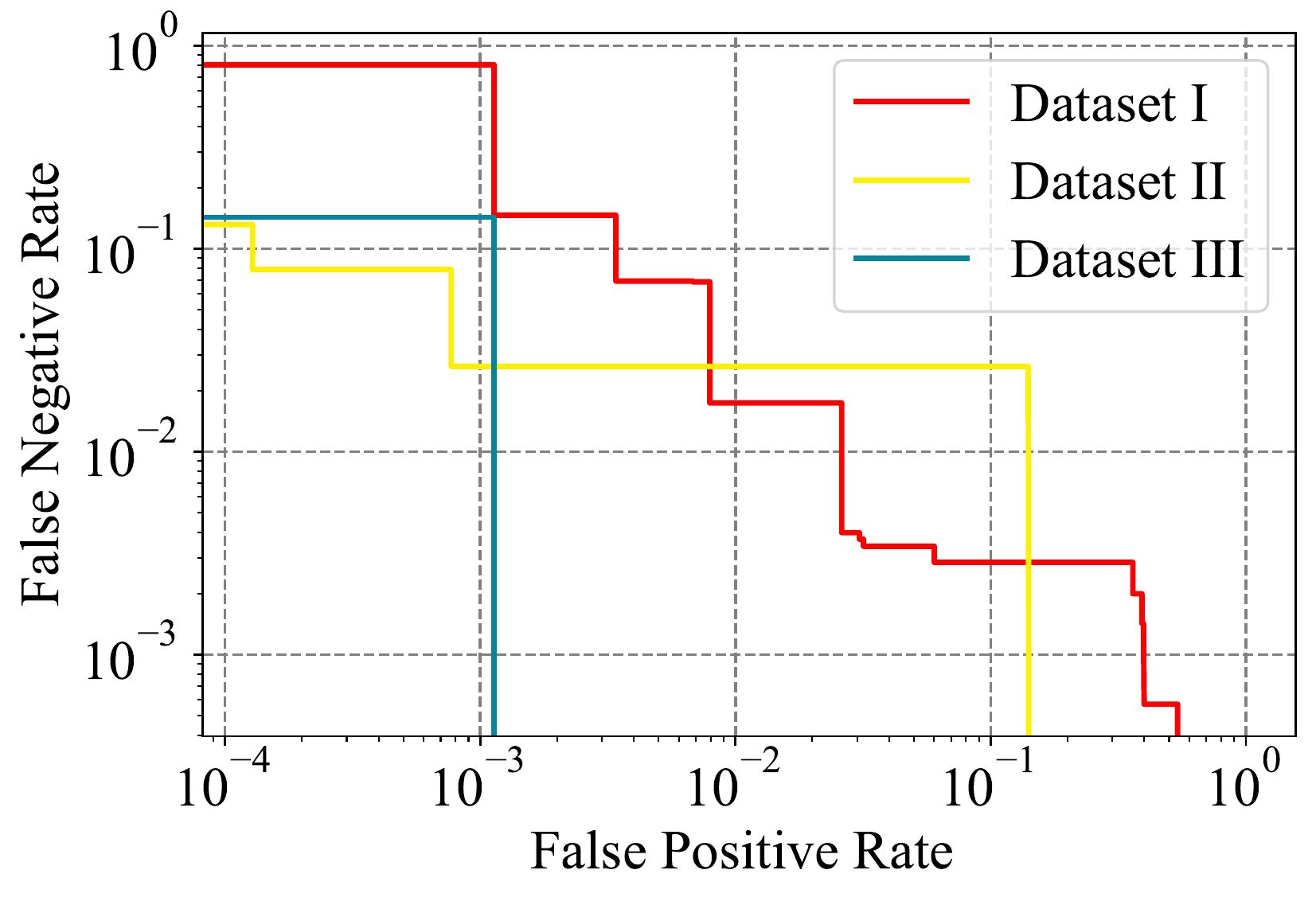}
}
\subfigure[Max Level $L$ = 10 ]{
\label{Fig_det_10}
\includegraphics[width=0.2\linewidth]{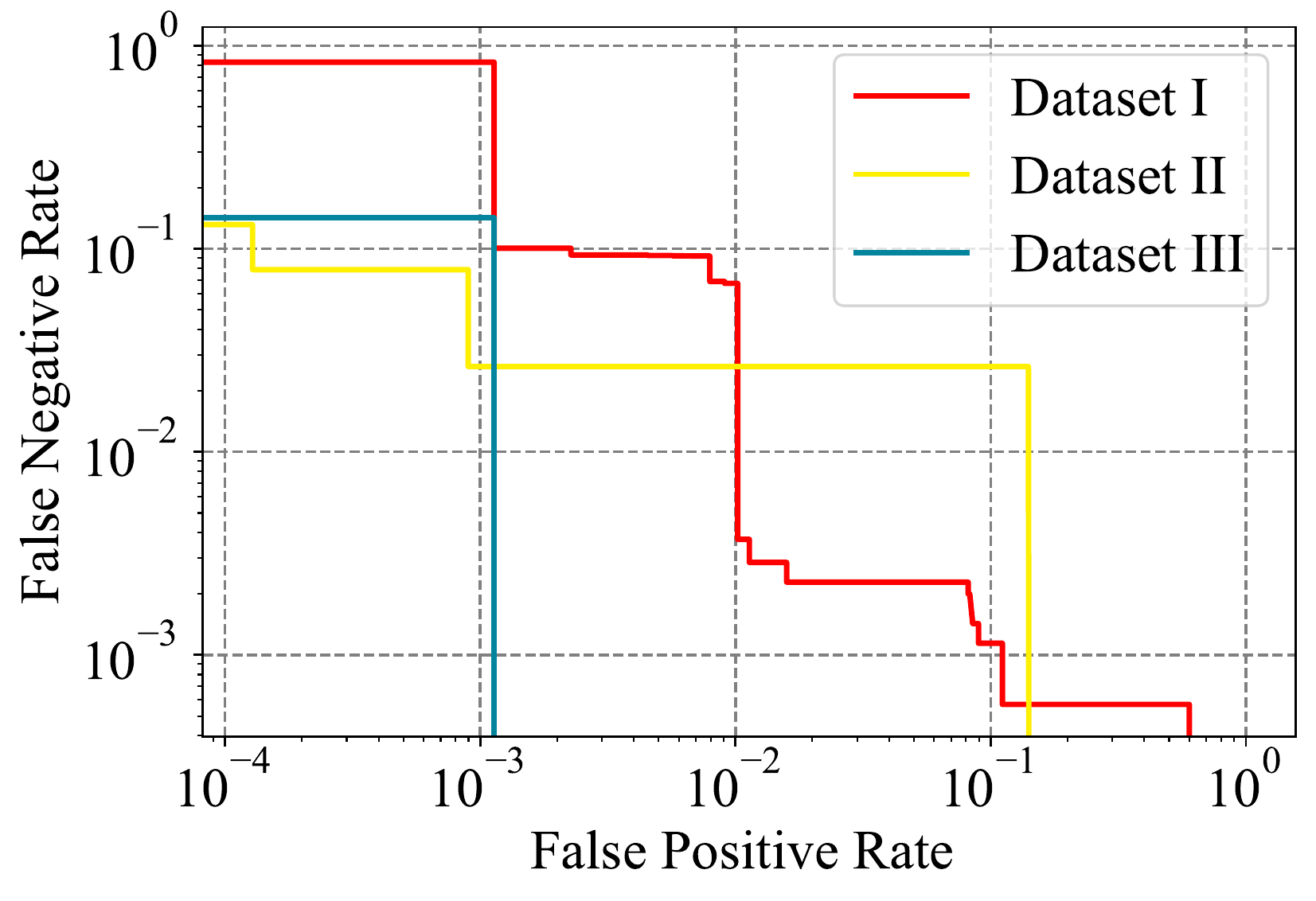}
}
\subfigure[Max Level $L$ = 15 ]{
\label{Fig_det_15}
\includegraphics[width=0.2\linewidth]{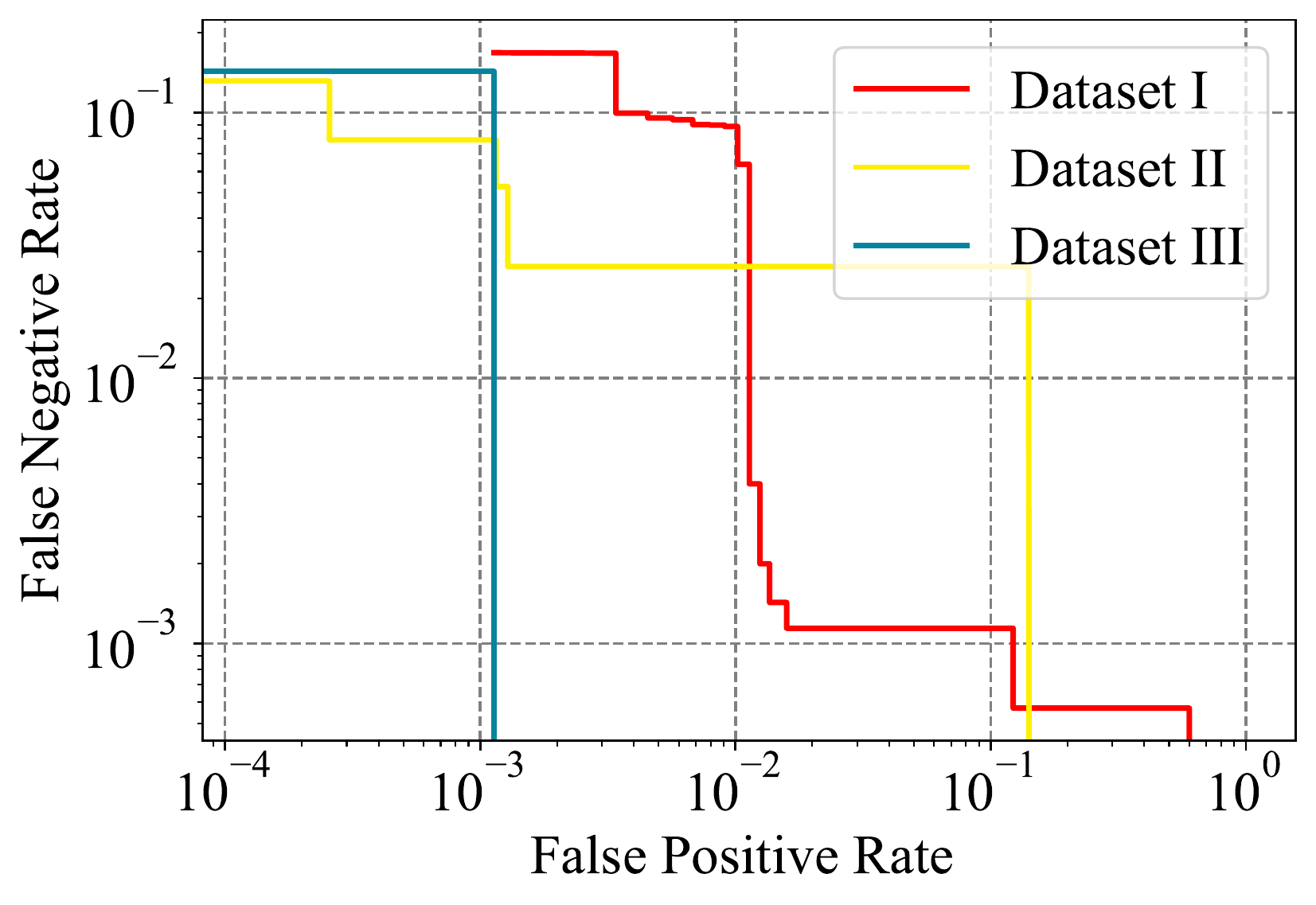}
}
\subfigure[Max Level $L$ = 20]{
\label{Fig_det_20}
\includegraphics[width=0.2\linewidth]{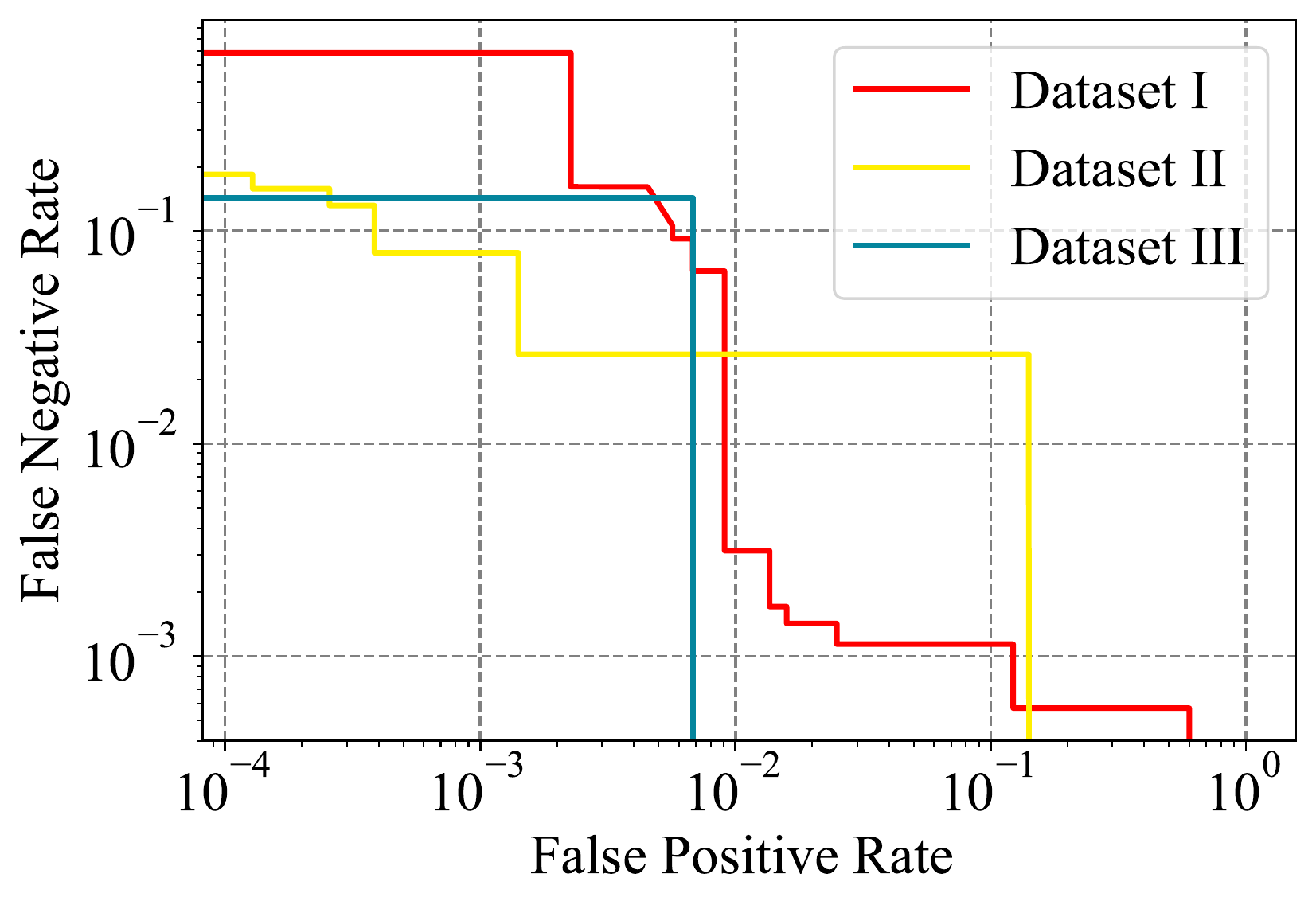}
}
\caption{DET curves with different Max Level $L$ and threshold $\theta$.} 
\label{Fig_det_threshold}
\end{figure*}
\subsection{Case Study}
To illustrate the advantage of MBTree, we provide a case study of \emph{koadic} RAT. 
The most interesting behavior of koadic is that it simulates the browser traffic by changing source port to disguise the C\&C traffic. Hence, session-based detection including most of the statistical-based methods and DirPiz-Seq can hardly capture the malicious patterns of koadic since the content is transferred separately. However, MBTree can well handle the disguised traffic by regarding all session traffic of a host as a whole part. 
And the similarity-based matching strategy can accurately identify the infected hosts by synthesizing the path similarity and node similarity.

\subsection{Efficiency Evaluation} 
\begin{table}[h]
\centering
\footnotesize
\caption{Prediction time of different methods for per instance.}
\label{tab_time}
\begin{tabular}{c|c|c|c|c}
\hline
Methods  & MBTree    & CART      & RF         & GBDT-CIC  \\ \hline
Time (s) & $10^{-5}$ & $10^{-7}$ & $10^{-5}$  & $10^{-3}$ \\ \hline 
Methods  & CNN       & SAE       & DirPiz-Seq &           \\ \hline
Time (s) & $10^{-2}$ & $10^{-3}$ & $10^{-1}$  &           \\ \hline
\end{tabular}
\end{table}
As a comprehensive evaluation, we also record the prediction time of each method as an efficiency comparison. By analyzing the workflow of the matching mechanism, we find that the main cost of our approach comes from the intersection operation in calculating the similarity score. Hence, we design a mechanism to realize parallel computing of similarity scores in our implementation. Specifically, for each signature, we start a process to calculate the corresponding similarity score and then collect all similarity scores for further steps. With this mechanism, the detection time is two order of magnitudes lower than that of the ordinary. The optimized results are shown in \autoref{tab_time}.
Apparently, MBTree has the same or even better performance than most statistical-based methods.

\subsection{Parameter Tuning}
In this section, we take experiments on the hyperparameters of MBTree to analyze their influences. Totally, there are four parameters, Max Level $L$, Scores Ratio $\alpha, \beta$, and the Threshold $\theta$. As adopted in previous experiments, we continue to use the default settings as start, L = 10, alpha = 0.3, beta = 0.7 and threshold = 2048. Results are shown in Fig. 7-9.

\begin{figure*}[h!]
\centering
\subfigure[Unique degree of head nodes in signatures of dataset I]{
\label{fig_sig_rat_head_nodes}
\includegraphics[width=0.25\linewidth]{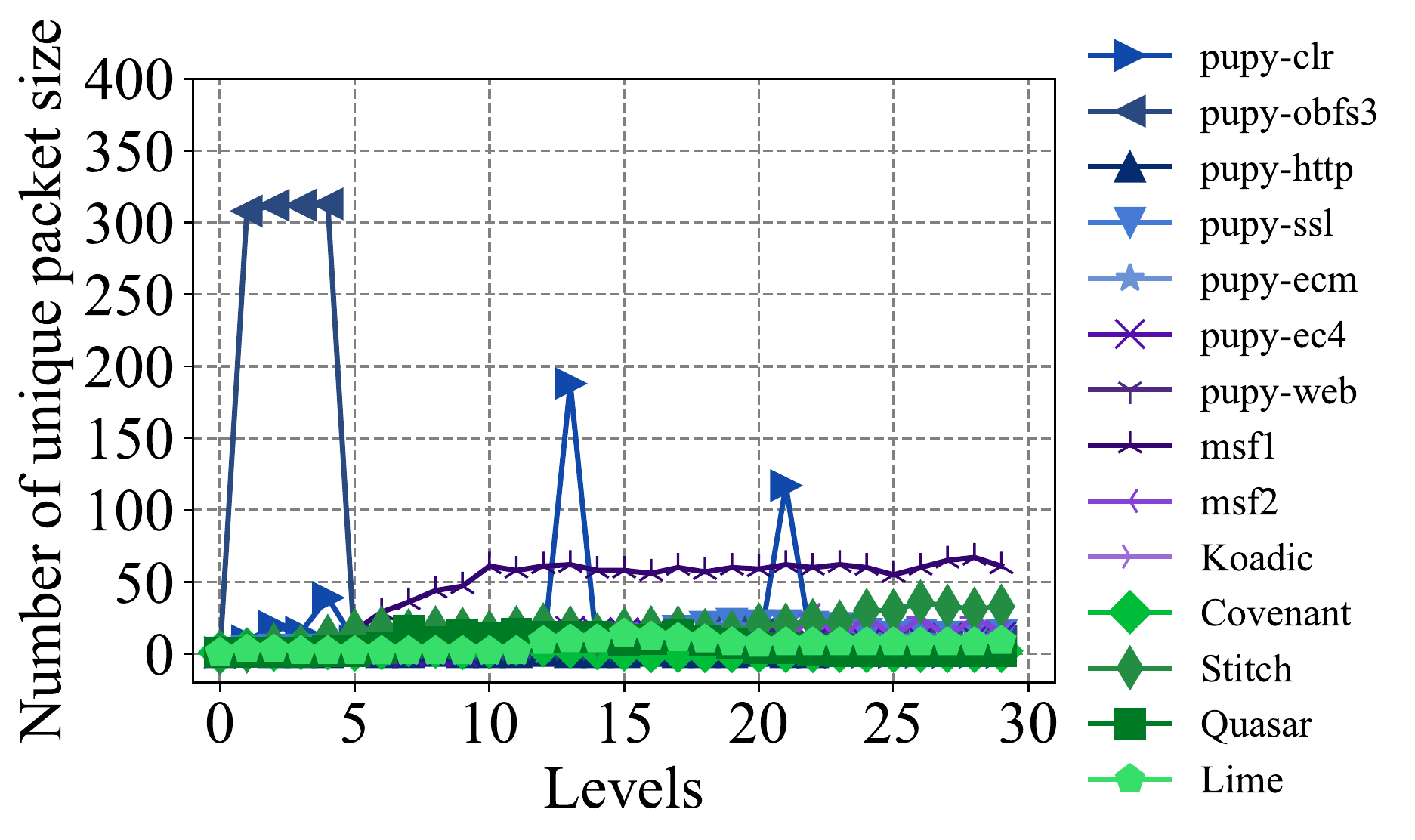}
}
\centering
\subfigure[Unique degree of tail nodes in signatures of dataset I]{
\label{fig_sig_rat_tail_nodes}
\includegraphics[width=0.25\linewidth]{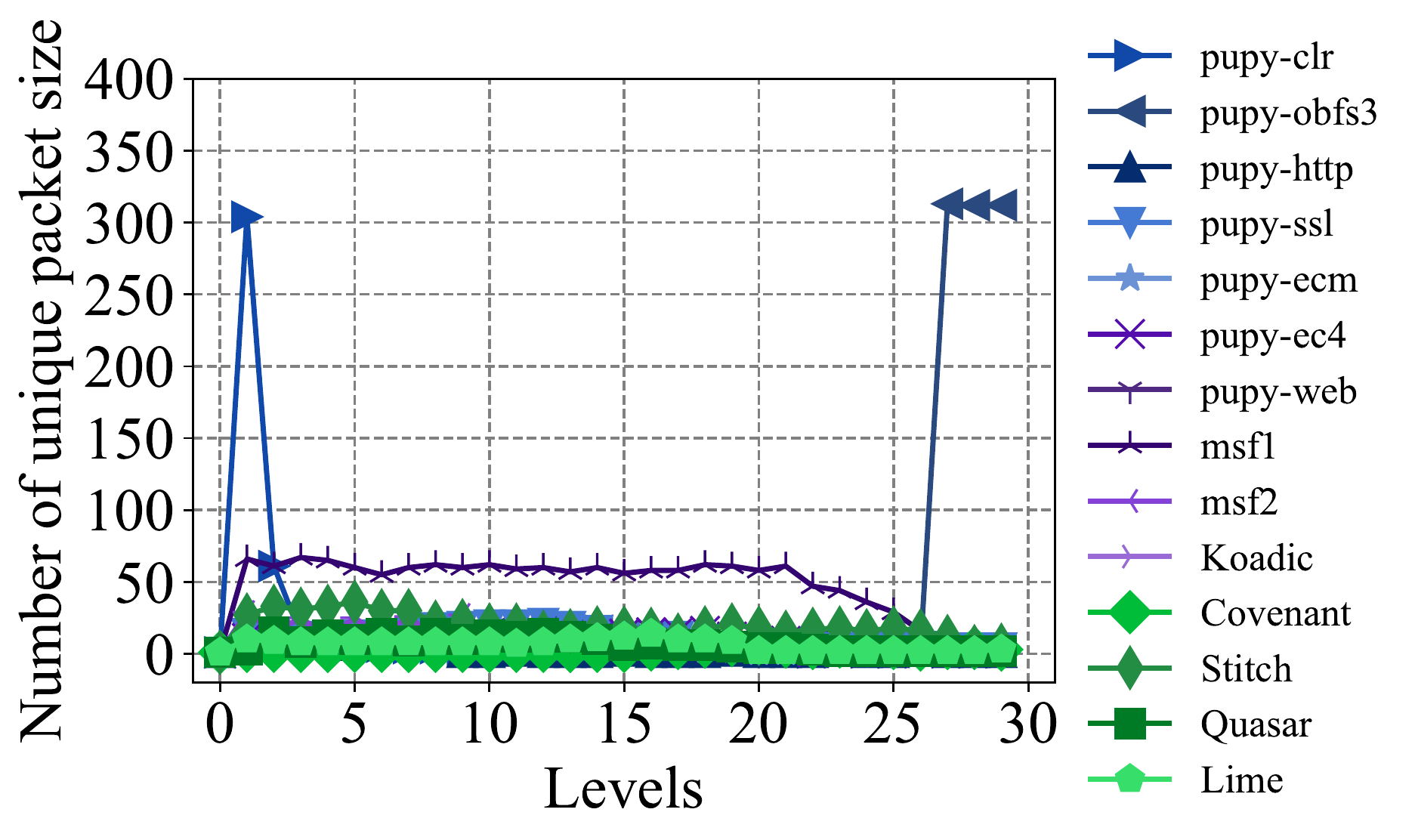}
}
\centering
\subfigure[Unique degree of head nodes in signatures of dataset II]{
\label{fig_sig_malware_head_nodes}
\includegraphics[width=0.2\linewidth]{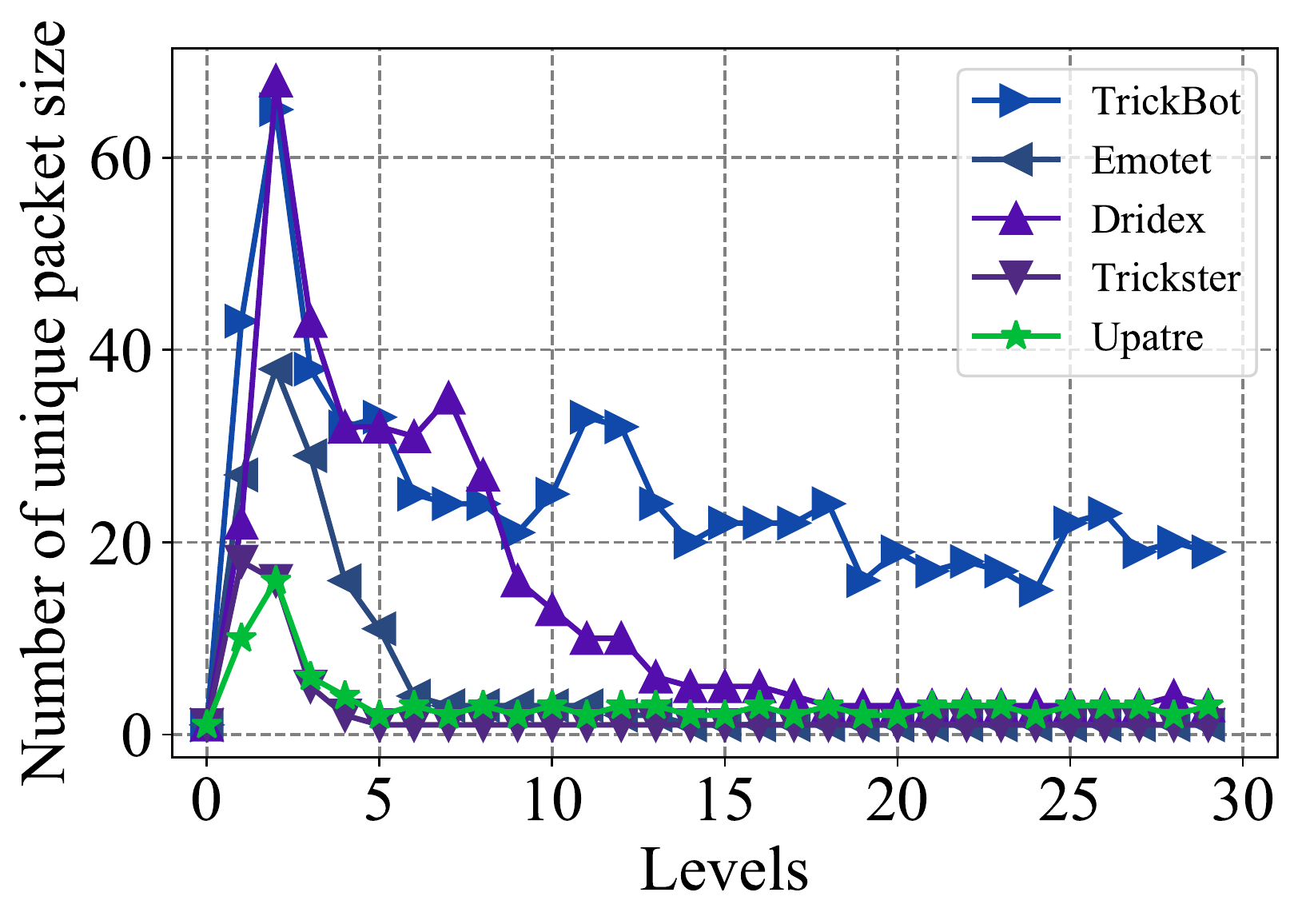}
}
\centering
\subfigure[Unique degree of tail nodes in signatures of dataset II]{
\label{fig_sig_malware_tail_nodes}
\includegraphics[width=0.2\linewidth]{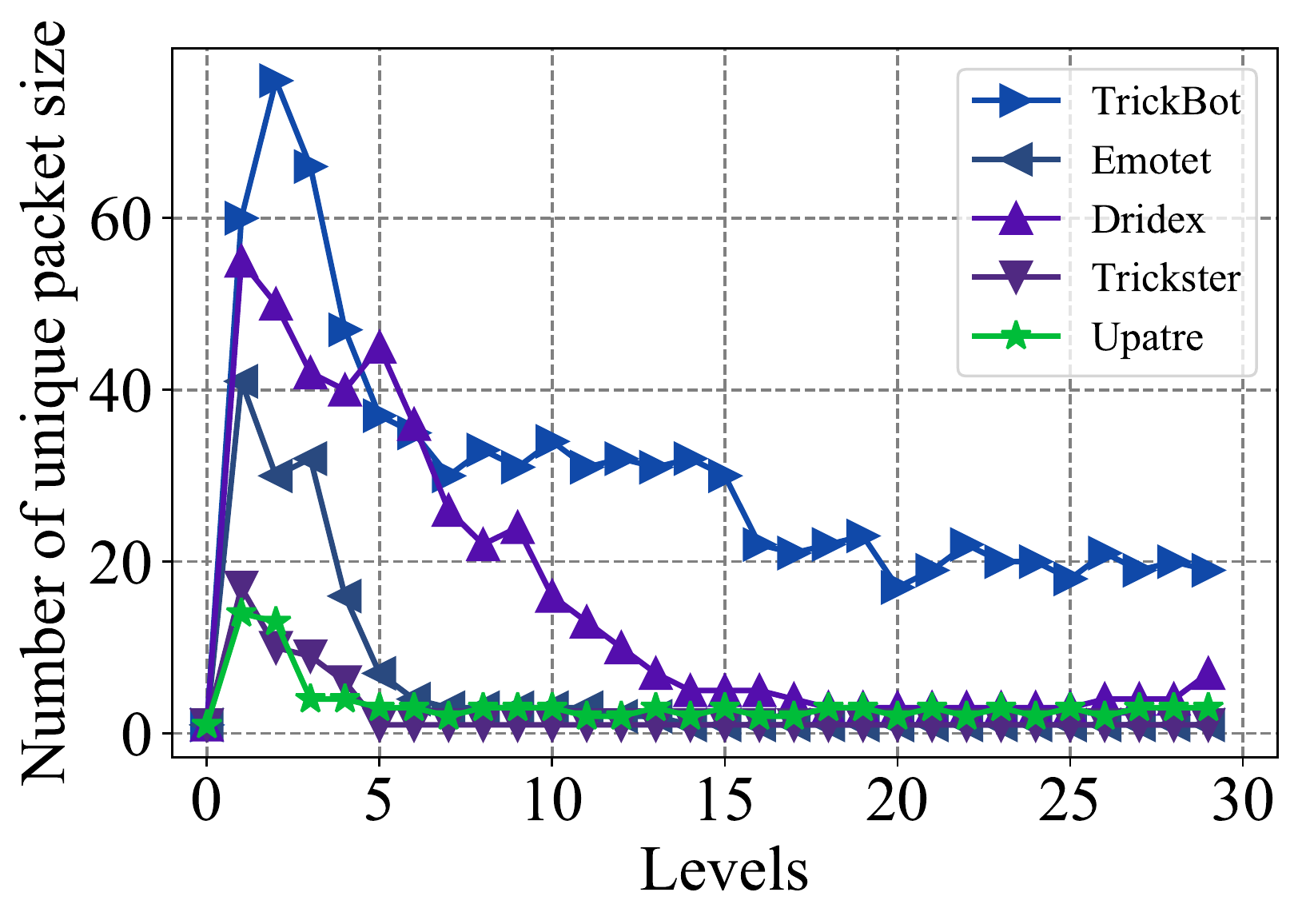}
}
\centering
\subfigure[Unique degree of head nodes in signatures of dataset III]{
\label{fig_sig_malware_head_nodes}
\includegraphics[width=0.2\linewidth]{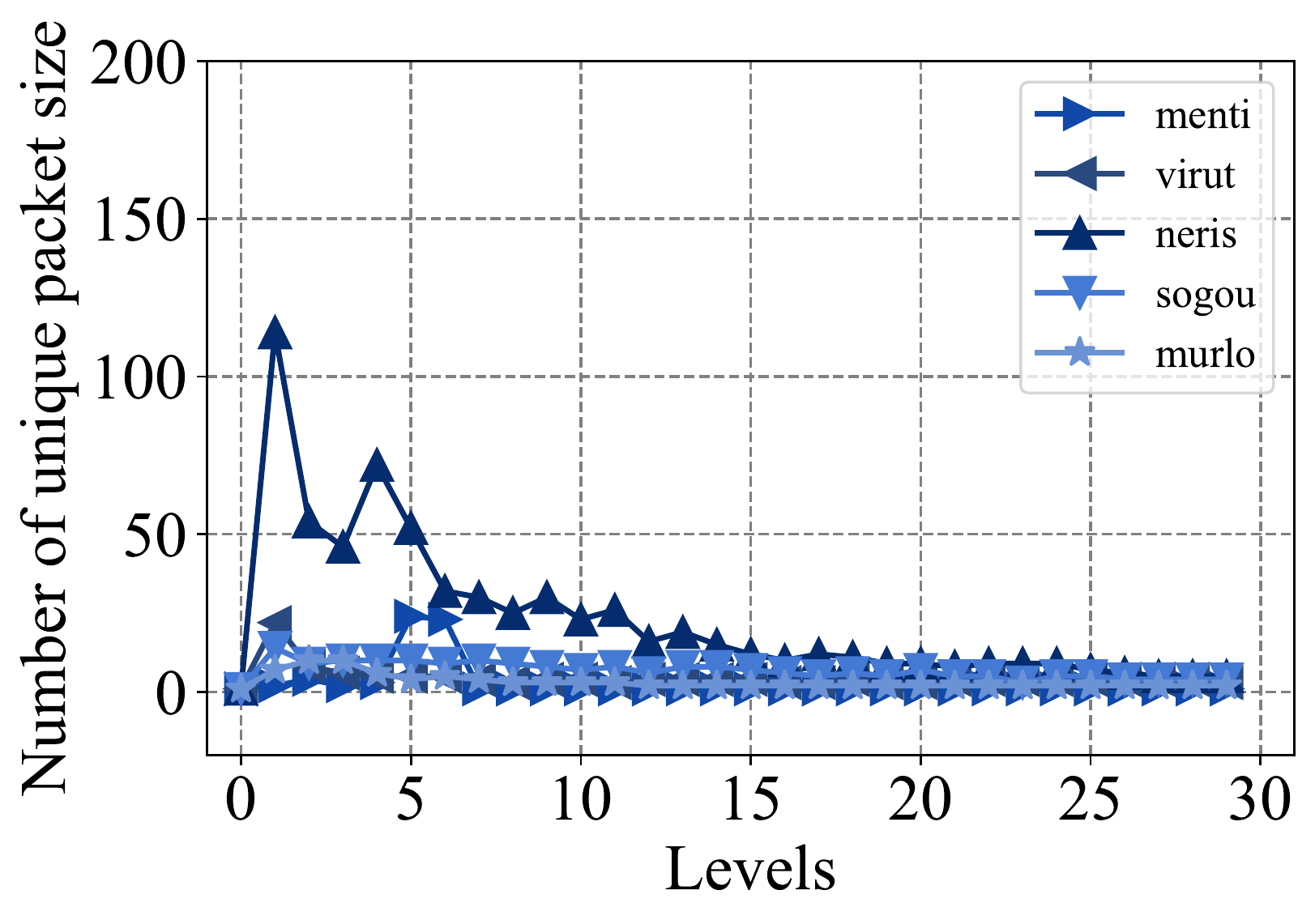}
}
\centering
\subfigure[Unique degree of tail nodes in signatures of dataset III]{
\label{fig_sig_malware_tail_nodes}
\includegraphics[width=0.2\linewidth]{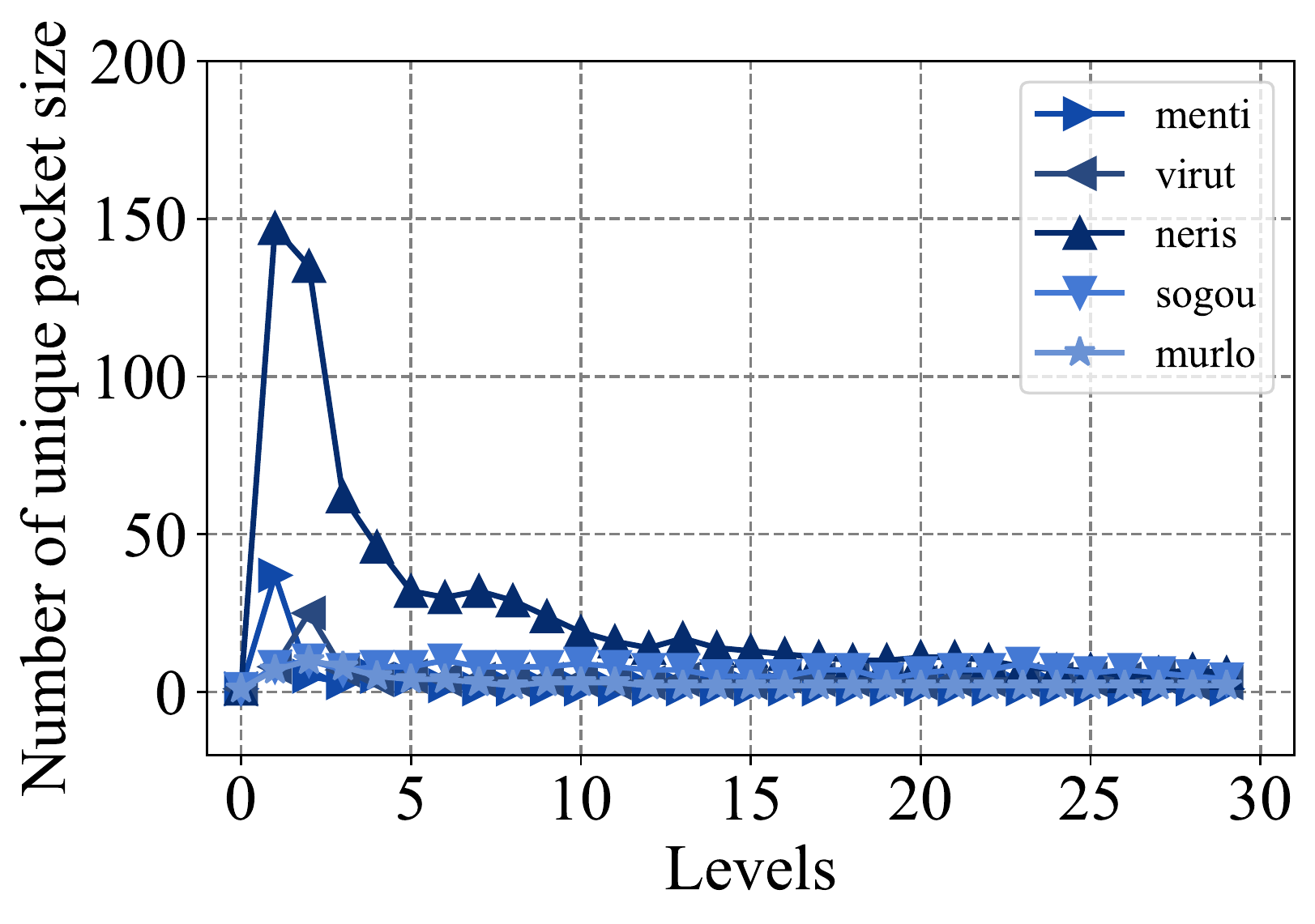}
}
\caption{Number of unique values of nodes and edges organized hierarchically in the generated MLTree signatures.
} 
\label{Fig_sigs}
\end{figure*}
\textbf{Path Score Ratio $\alpha$;}
The score ratio $\alpha$ determines the balance of the path score and node score. Specifically, $\alpha$ represents the ratio of the node score.
The results are shown in \autoref{Fig_alpha}. It can be noticed that the performance rise with $\alpha$ increasing at the initial on all datasets.
This can be mainly attributed to the existence of the dynamic DirPiz in the communication sequence, which truncates the CWP. 
Hence, we suggest that the $\alpha$ should be set lower than 0.75 under normal conditions to facilitate robust detection.

\textbf{Head Score Ratio $\beta$;} 
The parameter $\beta$ determines the balance of the head score and tail score. Specifically, $\beta$ represents the ratio of the head score.
The results are shown in \autoref{Fig_beta}. Slightly, the performances rise with $\beta$ increasing. This can be attributed to that the patterns of the handshake process are rather obvious. With the analysis of their traffic, it is obvious that the handshake process is sophisticated. The C\&C server sends fixed instructions (e.g., initialization) to the client after establishing the TCP connection.
While in the handwave process, it is rather simple. The server only sends one or two packets to notify the victim that the connection is closing, and then the rest work is handed over to the TCP level.
Hence, we suggest that the $\beta$ should be set greater than 0.5 without extra prior knowledge about samples.

\textbf{Max Level $\theta$;} The max level $\theta$ determines how deep should two MLTree be compared. Ideally, $L$ is supposed to be set as exactly the length of the automated handshake procedure. However, since the handshake process implementation varies from each other, it requires experiments on $L$ to determine the most appropriate value. Besides, for each $L$, we choose the corresponding theoretical best $\theta$. The results are shown in \autoref{Fig_maxlevel}. Obviously, based on the trend of the curves, the most appropriate value of $L$ should be lower than 15 to properly capture the automated behavior in most cases.

\textbf{Threshold $\theta$;}
The parameter $\theta$ determines the alarm level.
In this experiment, we observe the performance of different $\theta$ by setting different max levels $L$. Recall that $N_A$ is around 100. For each $L$, we sample 10 threshold points in the interval $[2^{L}, 2^{L+2}]$.
The \autoref{Fig_tunning_threshold} shows the F1 curve, and \autoref{Fig_det_threshold} shows the Detection Error Tradeoff (DET) curve.
It can be observed from \autoref{Fig_tunning_threshold} that the best performance is achieved around the points at [1,5], which covers the theoretical best threshold; thus this demonstrates the correctness of our theoretical analysis.

\subsection{Generated Signatures}
In this section, we analyze the generated signatures to inspect the behavior differences among different samples by counting the number of unique values at each level. The statistics are shown in \autoref{Fig_sigs}.
In this experiment, the max level $L$ is set as 30, which is a relatively deeper value for the handshake procedure.

First, it is apparent that the unique number of most signatures is less than 50 through different levels. 
However, pupy-obfs3 shows a different trend from most of the others. The DirPizs unique degree of this RAT is relatively high at the start of the communication because obfs3 achieves the traffic obfuscation in the handshake procedure by changing the packet size distribution \cite{obfs3}. However, even though obfuscation technology is adopted to randomize the packet size, MBTree can also effectively identify the traffic. This can be attributed to the limited range of DirPiz after randomization. Since the randomize strategy of obfs3 is applied as \emph{using random length of bytes to pad the rest of the packet}, the length of the padded DirPiz can be only in the interval $[raw\_content\_length, MTU]$ according to \cite{wang2015seeing, bitansky2018indistinguishability}.
Thus our designed node similarity can still cover the padded DirPiz sequences.
Second, comparing the head nodes with tail nodes,
it is apparent that most samples use the same packets to complete the handshake process. Thus, it can be deduced that the head patterns are more identical. This conclusion is also in accord with what we acquired in the experiment of $\beta$.
Third, rough automatic handshake length can be deduced based on the change points of the curves. For example, pupy-obfs3, msf-1, Dridex, TrickBot, and neris accomplish the handshake process through 5 message exchanges.

\section{Discussion}
In the previous section, the experiment results illustrate that even though the content is transferred through encrypted transport, MBTree still identifies malicious C\&C traffic. However, sophisticated attack strategies can still be taken by adversaries to paralyze MBTree. Here we discuss the strategies that can be used against MBTree.

\emph{Disguising Attack;} A potential attack strategy for MBTree is disguising malicious traffic as benign applications. When adopting this strategy, though the malicious traces can still be identified, it will \emph{pollute} MLTree signatures and result in a large number of false alarms to cripple the MBTree. However, in order to implement such a strategy, it requires the adversaries (i) to acquire which benign application is running on the victim machine; (ii) to keep on the update of the benign applications' behaviors, which is hard to achieve in practice.

\emph{Malformed Packet;}
Since MBTree relies on successful payload identification, the evading techniques exploiting the protocol stack parsing procedure can be used to evade MBTree\cite{wang2020symtcp, moon2019alembic}. For example, transfer content through RST packets. When adopting such a technique, the payload can be incorrectly reassembled by the man-in-the-middle MBTree, and the malicious patterns cannot be identified. However, implementing this strategy also requires extra protocol stack control to deal with the malformed packets correctly. Besides, traditional firewall or IDS can identified these malformed packets easily.

\emph{Obfuscation Attack;}
Although our experiment proves that MBTree can resist the obfuscation strategy to some extent, it will still lead to the invalidity of MBTree in the face of a highly targeted attack strategy.
A potential valid evading strategy is radically splitting the padded content into several packets with random packet size. In the case of such a strategy, not only will MBTree be polluted, resulting in high-level false alarms, but the sample can also evade the detection of MBTree with unseen DirPiz sequences. However, current RATs rarely attempt to hide their DirPiz identifications to our best known. Besides, to against this radical strategy, we suggest that the entropy analysis of the DirPiz sequences can be used. Since benign applications usually follow a specific procedure to complete the handshake, the entropy of their DirPiz sequences are relatively low; thus, it is abnormal if the entropy is exceedingly high in the case of only running a few applications on the machine.

\section{Conclusion}
In this paper, we present the MBTree, a novel signature-based approach that integrates DirPiz sequences as MLTree signatures with the similarity matching mechanism to detect encrypted RAT traffic. We evaluate MBTree against several C\&C traffic with comprehensive benign applications' traffic as background. The results show that MBTree can detect different malicious traffic in different environments with high-level accuracy.
Briefly, there are two directions in our future work to improve MBTree. First, we plan to improve the similarity score calculation, so as to detect malicious behaviors through different hierarchies. Second, inspired by \cite{bovenzi2020big}, we plan to improve the parallel computing ability by integrating the famous MapReduce framework \cite{dean2008mapreduce}. 

\appendix
\begin{table}[h]
\centering
\footnotesize
\caption{Selected open source RATs for generating malicious traffic. N1 denotes the session number in victim 1 as the training set. N2 denotes the session number in victim 2 as the testing set.}
\label{tab_rats}
\begin{tabular}{cccc}
\hline
Platfrom                 & Name                        & Transport                    & N1/N2\\ \hline
\multirow{10}{*}{Linux}  & \multirow{8}{*}{pupy}       & cleartext         & 500/500        \\
                         &                             & obfs3           & 300/500        \\
                         &                             & http            & 100/100        \\
                         &                             & SSL              & 100/100        \\
                         &                             & ECM              & 100/100        \\
                         &                             & RSA              & 500/500        \\
                         &                             & EC4+ECPV+RC4     & 100/100        \\
                         &                             & websock+RSA+AES   & 100/100        \\
                         & \multirow{2}{*}{Metasploit} & staged meterpreter & 200/500        \\
                         &                             & stageless meterpreter& 400/500        \\ \hline
\multirow{5}{*}{Windows} & koadic                      & TLS               & 150/300        \\
                         & Covenant                    & EKE+SSL           & 150/300        \\
                         & Stitch                      & AES                & 250/200        \\
                         & QuasarRAT                   & TLS                 & 100/100        \\
                         & Lime-RAT                    & AES                  & 100/100        \\ \hline
\end{tabular}
\end{table}

\begin{table}[h]
\footnotesize
\centering
\caption{An overview of the WT traffic.}
\label{tab_trojans}
\begin{tabular}{cccc}
\hline
Name      & Time           & Session Num & Host Num \\ \hline
TrickBot  & 2015.3-2018.4  & 152680   & 7        \\
Emotet    & 2017.6         & 344850   & 9        \\
Dridex    & 2017.3-2017.4  & 106427   & 13       \\
Trickster & 2017.6-2018.1  & 67301    & 3        \\
Upatre    & 2015.10-2016.5 & 114921   & 2        \\ \hline
\end{tabular}
\end{table}

\begin{table}[h]
\footnotesize
\centering
\caption{An overview of cleaned CTU-13 traffic.}
\label{tab_ctu13}
\begin{tabular}{cccc}
\hline
Name  & Session Num & Host Num \\ \hline
Neris & 2771        & 4        \\
Murlo & 1070        & 2        \\
Menti & 198         & 2        \\
Virut & 58          & 2        \\
Sogou & 40          & 3        \\ \hline
\end{tabular}
\end{table}

\bibliographystyle{unsrt}
\bibliography{thesis}

\end{document}